\documentclass[eqsecnum,showpacs,showkeys,twocolumn,superscriptaddress,aps,pre]{revtex4}
\usepackage{graphicx}
\usepackage{amsmath,amsfonts,amsthm}
\usepackage{color}

\begin{document}

\title{Identification of  a polymer growth process with an equilibrium multi-critical collapse phase transition: the meeting point of swollen, collapsed and crystalline polymers}

\author{Jason Doukas} \email{jasonad@yukawa.kyoto-u.ac.jp}
\altaffiliation[Current Address: ]{Yukawa Institute for Theoretical Physics, Kyoto University,
  606-8502, Japan}
\affiliation{Department of Mathematics and Statistics, The University
  of Melbourne, 3010, Australia}
\author{Aleksander L.\ Owczarek} \email{a.owczarek@ms.unimelb.edu.au}
\affiliation{Department of Mathematics and Statistics, The University
  of Melbourne, 3010, Australia}
\author{Thomas Prellberg} \email{t.prellberg@qmul.ac.uk}
\affiliation{School of Mathematical Sciences, Queen Mary University
  of London, Mile End Road, London E1 4NS, United Kingdom}

\begin{abstract}
  We have investigated a polymer growth process on the triangular lattice 
  where the  configurations produced are self-avoiding trails. We show that the scaling behaviour of this process is similar to the analogous process on the square lattice. However, while the square lattice process maps to the collapse transition of the canonical interacting self-avoiding trail model (ISAT) on that lattice, the process on the triangular lattice model does not map to the canonical equilibrium model. On the other hand, we show that the collapse transition of the canonical ISAT model on the triangular lattice behaves in a way reminiscent of the $\theta$-point of the interacting self-avoiding walk model (ISAW), which is the standard model of polymer collapse. This implies an unusual  lattice dependency of the ISAT collapse transition in two dimensions.
  
  By studying an extended ISAT model, we demonstrate that the growth process maps to a multi-critical point in a larger parameter space. In this extended parameter space the collapse phase transition may be either $\theta$-point-like (second-order) or first-order, and these two are separated by a multi-critical point. It is this multi-critical point to which the growth process maps. Furthermore, we provide evidence that in addition to the high-temperature gas-like swollen polymer phase (coil) and the low-temperature liquid drop-like collapse phase (globule) there is also a maximally dense crystal-like phase (crystal) at low temperatures dependent on the parameter values. The multi-critical point is the meeting point of these three phases. Our hypothesised phase diagram resolves the mystery of the seemingly  differing behaviours of the ISAW and ISAT models in two dimensions as well as the behaviour of the trail growth process. 
\end{abstract}

\pacs{05.50.+q, 05.70.fh, 61.41.+e}

\keywords{Interacting self-avoiding trails}

\maketitle

\section{Introduction}

Over the past 25 years various lattice models of a single
self-interacting polymer chain in (dilute) solution have been analysed in both two and
three dimensions. The fundamental physical phase
transition \cite{gennes1979a-a} that these models attempt to mimic 
is that of the collapse of a single polymer in a poor solvent as the
temperature is lowered. At high temperatures a polymer is swollen relative to a reference Gaussian state, while at low temperatures the polymer forms a liquid-drop-like globule \cite{gennes1979a-a,cloizeaux1990a-a}. There is a continuous phase transition expected between these two states, which is referred to as the $\theta$-point. One question that arises concerns the robustness  of the universality class of the
collapse transition. The standard theory
\cite{gennes1975a-a,stephen1975a-a,duplantier1982a-a} of the collapse
transition is based on the $n\rightarrow 0$ limit of the magnetic
tri-critical $\phi^4-\phi^6$ O($n$) field theory and related Edwards
model with two and three body forces
\cite{duplantier1986b-a,duplantier1987d-a}, which predicts an upper
critical dimension of three with subtle scaling behaviour in that
dimension. On the other hand recent studies \cite{bastolla1997a-a,krawczyk2009a-:a,krawczyk2010a-:a} of semi-flexible polymers indicate that collapse transition can be first-order, and a polymer crystal can exist at low temperatures in the presence of stiffness.

The canonical lattice model of the configurations of a polymer in solution has been the self-avoiding walk (SAW) 
where a random walk on a lattice  is not allowed to visit a lattice site more than once. Self-avoiding walks display the so-called excluded volume effect where they are swollen in size relative to unrestricted random walks of the same length: their size scales with a different characteristic exponent to that of the unrestricted random walk.  A common way \cite{gennes1979a-a} to model intra-polymer
interactions in such a walk is to assign an energy to each
non-consecutive pair of monomers lying on neighbouring lattice
sites. This is the interacting self-avoiding walk (ISAW) model, which is the standard lattice model of
polymer collapse using self-avoiding walks.

On the other hand self-avoiding trails \cite{shapir1984a-a}, which are lattice random walks that are not allowed to visit a lattice bond more than once also display the same excluded volume behaviour as SAW and physical polymers, with the same scaling exponents.  Moreover, a self-interacting self-avoiding trail model (ISAT), where multiply occupied sites are assigned an energy, has also displayed some of the characteristics of the polymer collapse described above \cite{lim1988a-a,owczarek1995a-:a,prellberg1995b-:a,owczarek2006c-:a}. 
However, analyses of both two- and three-dimensional self-interacting trails  
\cite{owczarek1995a-:a,prellberg1995b-:a,owczarek2006c-:a} indicate that
the collapse transition of the ISAT model is in a different universality class to that of self-interacting
self-avoiding walks (ISAW) in those respective dimensions.  There is no clear
understanding of why this is the case if true.  It is important to note that work in two dimensions has focused on the square lattice model only.

Some of the work \cite{owczarek1995a-:a,prellberg1995b-:a}  on the ISAT model has been via a growth process known as ``kinetic growth trails'' or ``smart kinetic trails'' that maps to one particular temperature of the equilibrium ISAT model on the lattices studied (it is important to note that these lattices are of coordination number four).   It was proposed in \cite{grassberger1996a-a} that the collapse transition associated with ``smart kinetic trails'' is first-order. Clear evidence was produced
in \cite{grassberger1996a-a} to demonstrate that there was a first-order
transition in three dimensions on the diamond lattice. On the other
hand no evidence of this could be found in two dimensions. In  fact, on the square lattice recent work \cite{owczarek2006c-:a} verified that the ISAT collapse transition on the square lattice is precisely that given by the growth process and is not first-order  or like the $\theta$-point of ISAW.

Another approach has been to analyse a generalised ensemble (rather than the finite length ensemble) of the model via transfer matrix \cite{foster2009a-a}: this has produced some intriguing results with different values of critical exponents being estimated. It was pointed out in \cite{foster2009a-a} that these results are not incompatible with those for the finite length ensemble. This may be an indication though that the point in question is multi-critical of some type.

In this paper we study the ISAT model and kinetic growth trails on the \emph{triangular} lattice, importantly a coordination number six lattice. We demonstrate that the kinetic growth model does not map to any temperature of the canonical ISAT model but rather to a point (we shall call this the \emph{kinetic growth point}) in the parameter space of a generalised model we call the Extended ISAT model (eISAT). As such we have studied this eISAT model and identify this kinetic growth point as a multi-critical collapse point. By studying the  eISAT model we build a picture of the collapse transition in the trail model of polymers that includes both the first-order transition suggested in \cite{grassberger1996a-a}  and the ISAW $\theta$-point. It may also prove useful in  explaining the transfer matrix results. 

\subsection{Review of previous results}
The collapse transition can be characterised via a change in the scaling of the size of the polymer with temperature. It is expected that some measure of the size, such as the radius of gyration or the mean squared distance of a monomer from the end points, $R_n^2(T)$, scales at fixed temperature as
\begin{equation}
R_n^2(T) \sim A n^{2\nu}
\end{equation}
with some exponent $\nu$. At high temperatures the polymer is swollen and in two dimensions it is accepted that $\nu=3/4$ \cite{nienhuis1982a-a}. At low temperatures the polymer becomes dense in space, though not space filling, and the exponent is $\nu=1/2$. The collapse phase transition is expected to take place at some temperature $T_t$. If the transition is second-order, the scaling at $T_t$ of the size is intermediate between the high and low temperature forms. In the thermodynamic limit there is expected to be a singularity in the free energy, which can be seen in its second derivative (the specific heat). Denoting the (intensive) finite length specific heat \emph{per monomer} by $c_n(T)$, the thermodynamic limit is given by the long length limit as
\begin{equation}
C(T) = \lim_{n\rightarrow\infty} c_n(T)\;.
\end{equation}
One expects that the singular part of the specific heat behaves as
\begin{equation}
C(T) \sim B |T_t -T|^{-\alpha}\;,
\end{equation}
where $\alpha<1$ for a second-order phase transition.
The singular part of the thermodynamic limit internal energy behaves as
\begin{equation}
U(T) \sim B |T_t -T|^{1-\alpha}\;,
\end{equation}
if the transition is second-order, and there is a jump in the internal energy if the transition is first-order (an effective value of $\alpha=1$).

 Moreover one expects crossover scaling forms \cite{brak1993a-:a} to apply around this temperature, so that
 \begin{equation}
 c_n(T) \sim n^{\alpha\phi} \; {\cal C}((T - T_t)n^\phi)
\end{equation}
with  $0<\phi < 1$ if the transition is second-order and 
  \begin{equation}
 c_n(T) \sim n \; {\cal C}((T - T_t)n)
\end{equation}
 if the transition is first-order.
 From \cite{brak1993a-:a} we point out  that the exponents $\alpha$ and $\phi$ are related via
\begin{equation}
2-\alpha = \frac{1}{\phi}\;.
\end{equation}

 \paragraph{$\theta$-point ISAW collapse}
The work of Duplantier and Saleur \cite{duplantier1987a-a} predicted the standard $\theta$-point behaviour in two dimensions, which has been subsequently verified \cite{prellberg1994a-:a}. It is expected that
\begin{equation}
\phi =3/7\approx 0.43 \quad \mbox{ and } \quad \alpha=-1/3\;. 
\end{equation}
Note that this implies that the specific heat does \emph{not} diverge at the transition. However, the third derivative of the free energy with respect to temperature will diverge with exponent $(1+\alpha)\phi=2/7$. At $T=T_t$ it is expected that
\begin{equation}
R_n^2(T) \sim A n^{8/7}\;. 
\end{equation}

 \paragraph{ISAT collapse on the square lattice}
Previous work \cite{owczarek2006c-:a}  on the square lattice has shown that  there is a collapse transition with a  strongly divergent specific heat, and the exponents have been estimated as 
\begin{equation}
\phi =0.84(3)\quad \mbox{ and } \quad \alpha=0.81(3)\;. 
\end{equation}
At $T=T_t$ it was  predicted \cite{owczarek1995a-:a} that
\begin{equation}
R_n^2(T) \sim A n\left(\log n\right)^2\;.
\end{equation}

\section{Trail growth on the triangular lattice}

\begin{figure}[ht!]
\begin{center}
\includegraphics[width=\columnwidth]{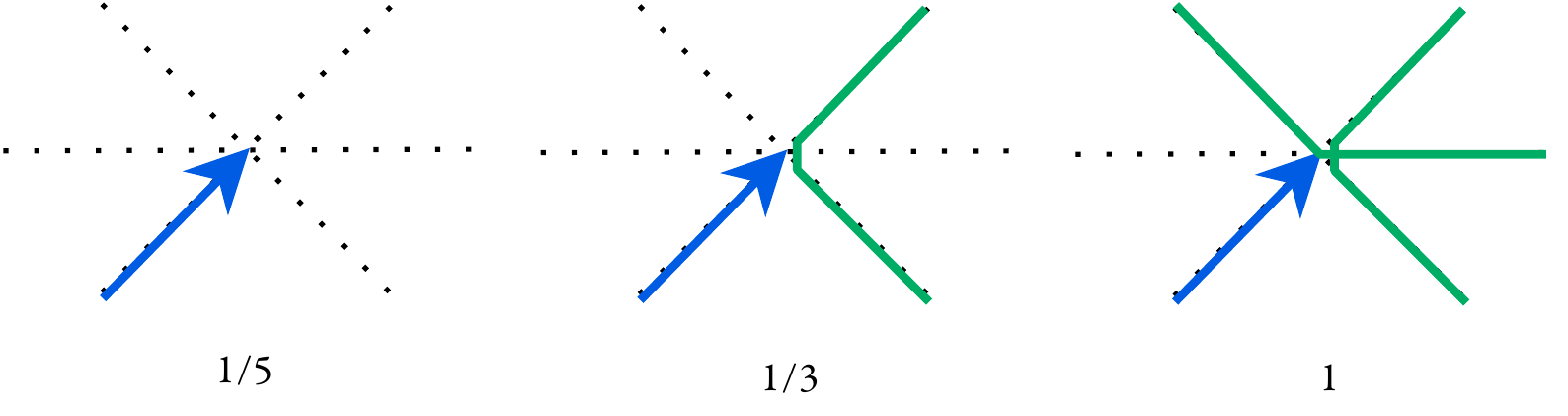}
\end{center}
\caption{This figure illustrates the growth process on the triangular lattice. Steps are created with different probabilities: the probabilities of the \emph{next} steps (not shown) are displayed underneath each case. Three general possibilities exist as the trail grows:  firstly, that the trails enters a site that is previously unoccupied --- in that case the five choices of next step are chosen at random with equal probability, that is, $1/5$; secondly,  that the trail enters a site that has been  previously visited once --- in that case  the three remaining choices of next step are chosen at random with equal probability, that is, $1/3$;  finally,  that the trail enters a site that has been  previously visited twice --- there is only one choice of continuation and this is made with probability $1$.}
\label{growthchoices} 
\end{figure}

\begin{figure}[ht!]
\begin{center}
\includegraphics[width=0.8\columnwidth]{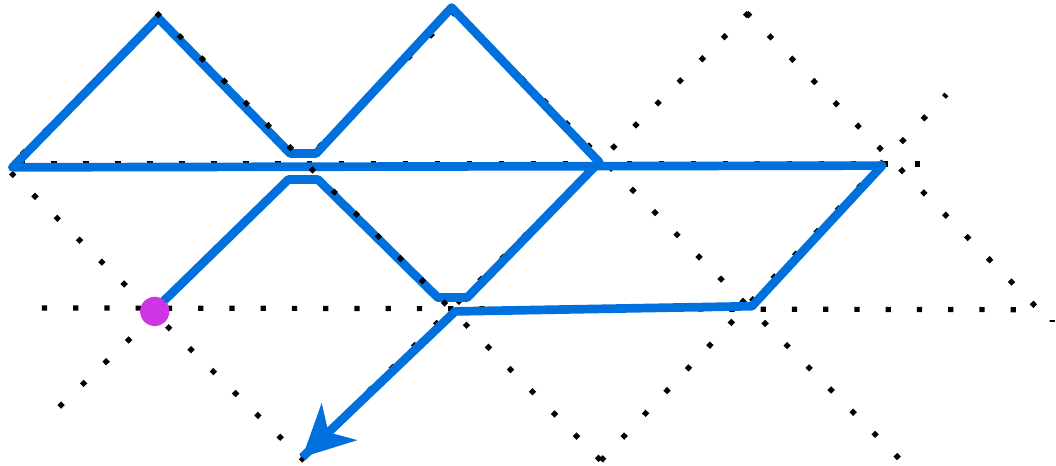}
\end{center}
\caption{An example of a trail with 13 steps on the triangular lattice. This trail has six singly visited sites, two doubly-visited sites and one triply-visited site. Note that the path may cross or touch at a visited site. This trail is produced by the growth process with probability 
$(\frac{1}{6})(\frac{1}{5})(\frac{1}{5})(\frac{1}{5})(\frac{1}{5})(\frac{1}{3})(\frac{1}{5})(\frac{1}{5})(1)(\frac{1}{3})(\frac{1}{5})(\frac{1}{5})(\frac{1}{3})$ --- the individual fractional probabilities are the step probabilities in order of creation of the steps.}
\label{trail-example} 
\end{figure}

Consider a stochastic process defined on the triangular lattice as follows:
Starting at an origin site, a lattice path is built up step-by-step by choosing between available continuing steps from unoccupied lattice bonds with equal probability.  The configuration produced is a \emph{self-avoiding trail} or \emph{trail} for short, where sites may be visited multiple times but bonds of the lattice are either visited once or not at all. 
If we do not consider the original occupation of the origin as a visit, then each site may be visited up to three times on the triangular lattice.
Note that when the process revisits the origin, the growth rule is slightly altered: the process may choose the direction of the first step as one of its options equally with the unoccupied lattice bonds.
If it does this, a loop is formed and the growth process terminates. 
Apart from the initial step, where there are six available steps, the number of available steps is therefore five minus twice the number of previous visits. 
A picture of these three situations is provided in Figure~\ref{growthchoices}, along with the probabilities of the \emph{next} step that is added. 
In Figure~\ref{trail-example} an example of a configuration produced by the process is illustrated. Apart from the origin, each singly visited site  contributes a factor $(\frac{1}{5})$ to the overall probability, while each twice visited site contributes a factor $(\frac{1}{5})(\frac{1}{3})$ and each triply-visited site contributes a factor  $(\frac{1}{5})(\frac{1}{3}) (1)$. 
This model is the triangular lattice version of the growth model considered previously \cite{owczarek1995a-:a}.

We will not count the initial occupation of the origin as a visit. If we denote the number of steps of the trail as $n$, and  the numbers of singly, doubly and triply-visited site as $m_1$, $m_2$ and $m_3$, respectively, then these satisfy
\begin{equation}
n = m_1+2m_2+3m_3\;.
\label{m-constraint}
\end{equation}

The probability of a configuration, $\varphi_n$, of $n$ steps is denoted $p_G(\varphi_n)$, and we have for each $n$ that 
\begin{equation}
\sum_{\varphi_n} p_G(\varphi_n) =P_n\;,
\end{equation}
where $P_n$ is the probability that the growth process reaches length $n$.
Let us define the expectation values of the number of singly, doubly and triply-visited sites per unit length $e_j(n)$ for the growth process conditioned on the process making it to length $n$ respectively as
\begin{equation}
e_j(n) = \frac{\langle m_j\rangle}{n} = \frac{1}{nP_n} \sum_{\varphi_n} p_G(\varphi_n)\;  m_j(\varphi_n)\;,
\end{equation}
and the fluctuations in these as $f_j(n)$ with
\begin{equation}
f_j(n) = \frac{\langle m^2_j\rangle  - \langle m_j\rangle^2}{n} \;.
\end{equation}
Note that from Equation~(\ref{m-constraint}) we have 
\begin{equation}
e_1(n)+2e_2(n)+3e_3(n) =1\;.
\label{e-constraint}
\end{equation}

In the square lattice case \cite{meirovitch1989d-a,bradley1990a-a} this stochastic model can be mapped to a specific temperature of an equilibrium model, and moreover \cite{owczarek1995a-:a,owczarek2006c-:a} it was shown that this temperature was a critical point of the equilibrium model. There, the exponents $\alpha$ and $\phi$ defined in the introduction can be related to the behaviour of $e_2$ and $f_2$.  Assuming a similar form of  critical behaviour is found in the triangular lattice model we would expect that
\begin{equation}
e_2(n) \sim E_2 - \frac{a_2}{n^{(1-\alpha)\phi}}
\label{e2-scale}
\end{equation}
and 
\begin{equation}
e_3(n) \sim E_3 - \frac{a_3}{n^{(1-\alpha)\phi}}\;.
\label{e3-scale}
\end{equation}
Also
\begin{equation}
f_2(n) \sim b_2 \; n^{\alpha\phi} +F_2
\label{f2-scale}
\end{equation}
and 
\begin{equation}
f_3(n) \sim b_3 \; n^{\alpha\phi} +F_3\;,
\label{f3-scale}
\end{equation}
where either the $n^{\alpha\phi}$ or the constant term dominates dependent on whether $\alpha$ is positive or negative.

\begin{figure}[ht!]
\begin{center}
\includegraphics[width=0.9\columnwidth]{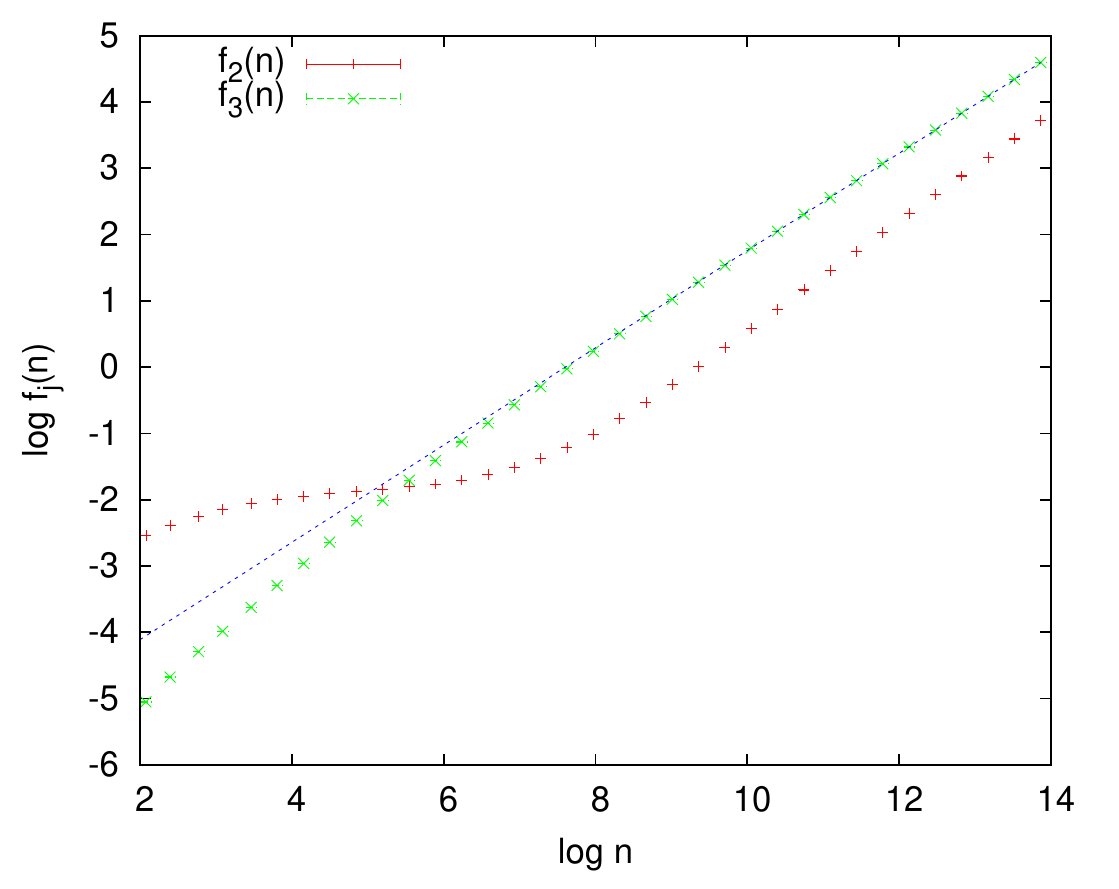}
\end{center}
\caption{Plot of the fluctuations in the number of doubly and triply-visited sites $f_2(n)$ and $f_3(n)$ against $n$. The slope gives us an estimate of $\alpha\phi$. }
\label{triplefluctuations} 
\end{figure}

\begin{figure}[ht!]
\begin{center}
\includegraphics[width=0.9\columnwidth]{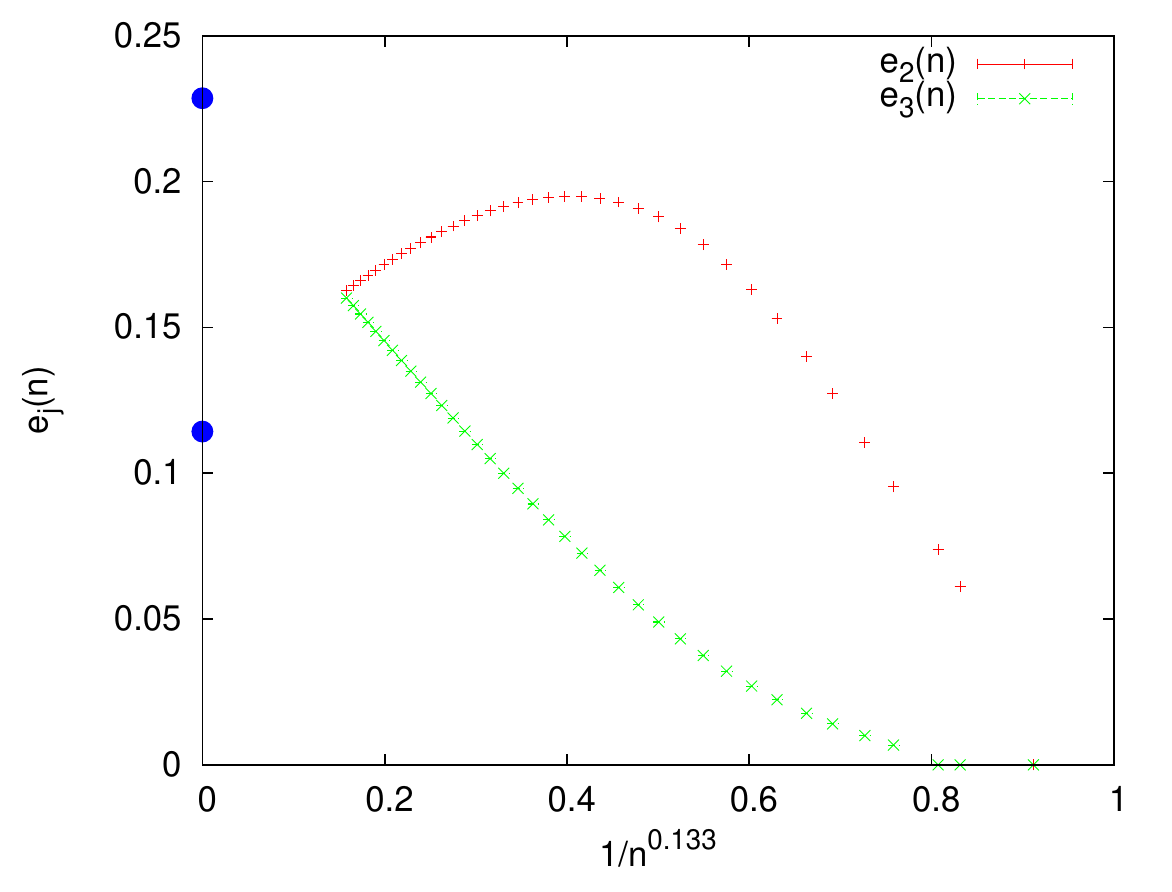}
\end{center}
\caption{Plot of the average numbers of doubly, $e_2(n)$, and triply, $e_3(n)$, visited sites against $1/n^{0.133}$. Their asymptotic values, $E_2=4/35\approx 0.114$ and $E_3=8/35\approx0.228$ respectively are marked as filled circles. }
\label{growth-sites} 
\end{figure}

We have simulated this growth process for lengths up to $n=2^{20}=1,048,576$ producing $7,120,000$ samples calculating estimates for $e_2(n)$, $e_3(n)$, $f_2(n)$, and $f_3(n)$. In Figure~\ref{triplefluctuations} we plot $f_2(n)$ and $f_3(n)$ against $n$: we immediately note that both quantities diverge. 
Taking the better behaved data for $f_3(n)$, we have estimated
\begin{equation}
\alpha\phi=0.734(6)\;.
\end{equation}
We can therefore deduce the estimates 
\begin{equation}
\alpha = 0.847(3) \quad \mbox{ and } \quad \phi = 0.867(3)\;.
\end{equation}
Importantly this is consistent with the estimates found in \cite{owczarek1995a-:a} for the growth process on the square lattice where $\phi$ was estimated to be $0.88(7)$, subsequently confirmed independently and directly for the equilibrium model in \cite{owczarek2006c-:a} where $\phi=0.84(3)$ was found.
Our estimate of $\alpha\phi=0.734(6)$ implies that the exponent involved in the scaling of the $e_j$, namely $(1-\alpha)\phi$, takes the value $0.133(4)$. 
In Figure~\ref{growth-sites} we plot $e_2(n)$ and $e_3(n)$ against $1/n^{0.133}$. We immediately see the strong corrections to the scaling forms still apparent even at the long lengths we have simulated. The maximum in $e_2(n)$ at $n\approx 1000$ mirrors the behaviour found in $f_2(n)$ in Figure~\ref{triplefluctuations}. 

In Figure~\ref{growth-sites}  we have also marked the asymptotic values $E_2$ and $E_3$ which can be deduced from the following argument.
Consider the case when a trail has formed a large $n$-step loop which occupies $m$ lattice sites.
Any site of this loop could have been the starting point. In order for this site to be visited
only once, the loop must have closed at the first return visit, which occurs with a probability
of $(\frac15)$. In order for this site to be visited twice, the loop must have closed at the second
return visit, which occurs with a probability $(1-\frac15)(\frac13)$. Finally, for this
site to be visited three times, the loop closes at the third return visit, which occurs with a
probability $(1-\frac15)(1-\frac13)$. Therefore we find for large loops the asymptotic
values $m_1/m=1/5$, $m_2/m=4/15$, and $m_3/m=8/15$. Using (\ref{m-constraint}), we obtain
\begin{equation}
e_1(n)\to\frac3{35}\;,\quad e_2(n)\to\frac4{35}\;,\quad\text{and}\quad e_3(n)\to\frac8{35}
\end{equation}
as $n\to\infty$,
whence we identify $E_2=4/35\approx0.114$ and $E_3=8/35\approx0.228$.

\section{Canonical ISAT model on the triangular lattice}
The canonical model \cite{lim1988a-a} of self-interacting trails (ISAT) on the triangular  lattice is
defined as follows. Consider all different bond-avoiding paths (open trails and loops)
$\varphi_n$ of length $n$ that can be formed on the triangular lattice
with one end fixed at a particular site (the set $\Omega_n$). Associate
an energy $-\varepsilon$ with each doubly-visited site and an energy $-2\varepsilon$ with each triply-visited site. 
For each configuration $\varphi_n$ count the number $m_2(\varphi_n)$ of
doubly-visited sites and $m_3(\varphi_n)$ of
triply-visited sites  of the lattice and give that configuration a
Boltzmann weight $\omega^{m_2+2m_3}$, where $\omega=\exp(\beta\varepsilon)$. The
partition function of the ISAT model is then given by
\begin{equation}
Z^{(2)}_n(\omega) = \sum_{\varphi_n\in \Omega_n} \omega^{m_2(\varphi_n)+2m_3(\varphi_n)} \; .
\end{equation}
We use the superscript $(2)$ to denote the canonical model. The choice of this notation will become clear in the next section.
The average of any quantity $Q$ over the
ensemble set $\Omega_n$ of allowed paths of length $n$ is given
generically by
\begin{equation}
\langle Q \rangle_n (\omega)= \frac{\sum_{\varphi_n\in\Omega_n} Q(\varphi_n) \omega^{m_2(\varphi_n)+2m_3(\varphi_n)}}{Z^{(2)}_n(\omega)}\; .
\label{canonical-isat-average}
\end{equation}
The reduced free energy $\kappa^{(2)}_n(\omega)$ per step  is given by
\begin{equation}
\kappa^{(2)}_n(\omega) = \frac{1}{n} \log Z^{(2)}_n(\omega)\;,
\end{equation}
and the internal energy $u^{(2)}_n(\omega)$ and specific heat $c^{(2)}_n(\omega)$ can be found as the first and second derivatives of the free energy with respect to $\beta$. We will also require the third derivative $t^{(2)}_n(\omega)$. However, these can all be found from the expected values of moments of $m_2$ and $m_3$. The internal energy (setting $\varepsilon=1$) is given by
\begin{equation}
u^{(2)}_n(\omega) =  \frac{\langle m_2 +2 m_3 \rangle}{n}
 \end{equation}
 and the (reduced) specific heat is given by
 \begin{equation}
 c^{(2)}_n(\omega)= \frac{\langle (m_2 +2 m_3)^2\rangle  - \langle (m_2 +2 m_3) \rangle^2}{n} \; .
 \end{equation}

\begin{figure}[ht!]
\begin{center}
\includegraphics[width=0.9\columnwidth]{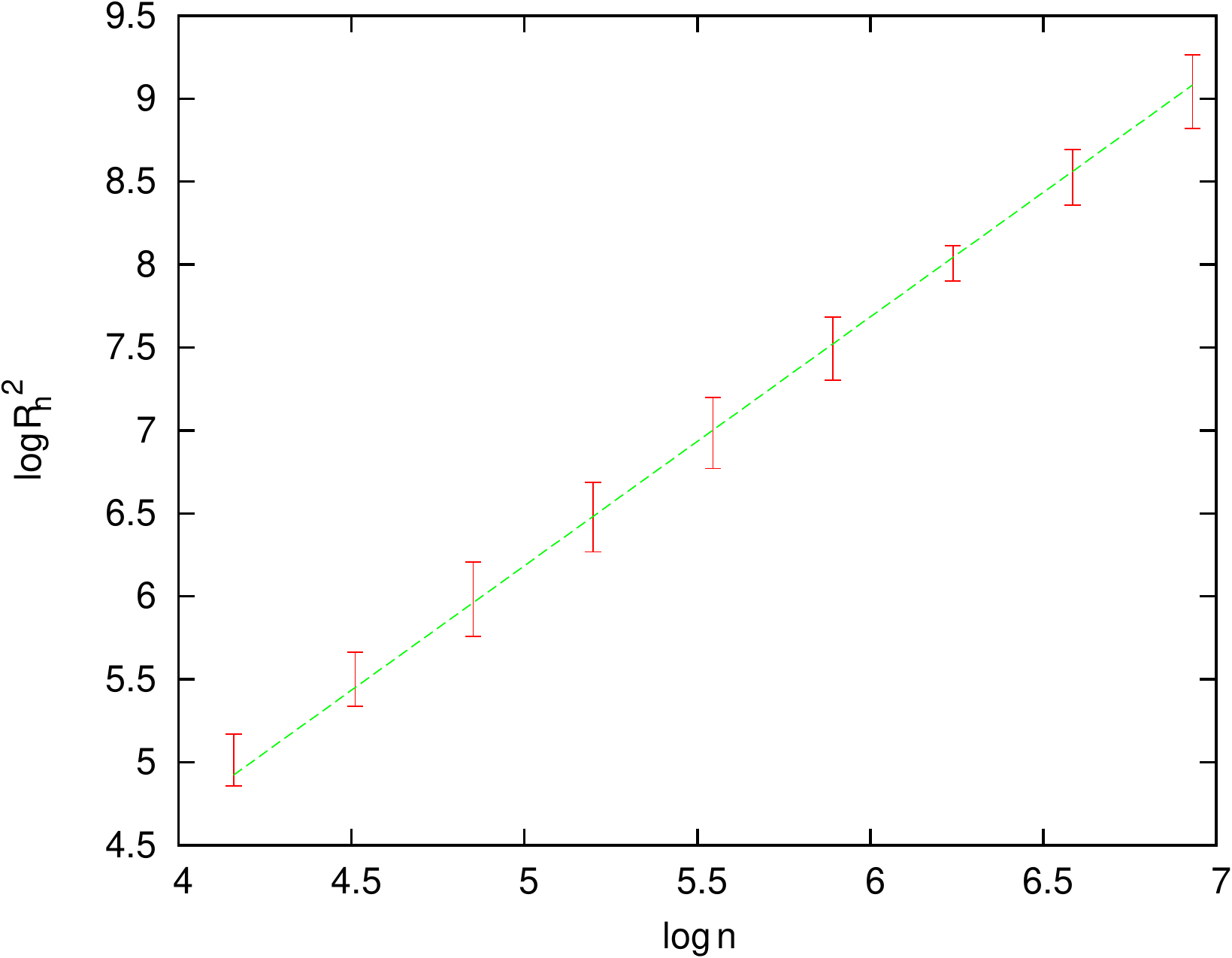}
\end{center}
\caption{A plot of $\log R_n^2$ against $\log n$ at $(\omega_2,\omega_3)=(1.5,2.25)$ in the swollen phase showing a slope of $2\nu=3/2$. }
\label{logR2vslogN} 
\end{figure}

We have simulated the canonical ISAT model using FlatPERM \cite{prellberg2004a-a} for lengths up to $1024$, generating $1.6\times10^7$ samples at that length. 
First we consider a high temperature point at $\omega=1.5$ and show that the scaling of the size of the trail is in accord with $\nu=3/4$ and the standard swollen phase expectations: in Figure~\ref{logR2vslogN} we display a plot of $\log R_n^2$ against $\log n$. 

\begin{figure}[ht!]
\begin{center}
\includegraphics[width=0.9\columnwidth]{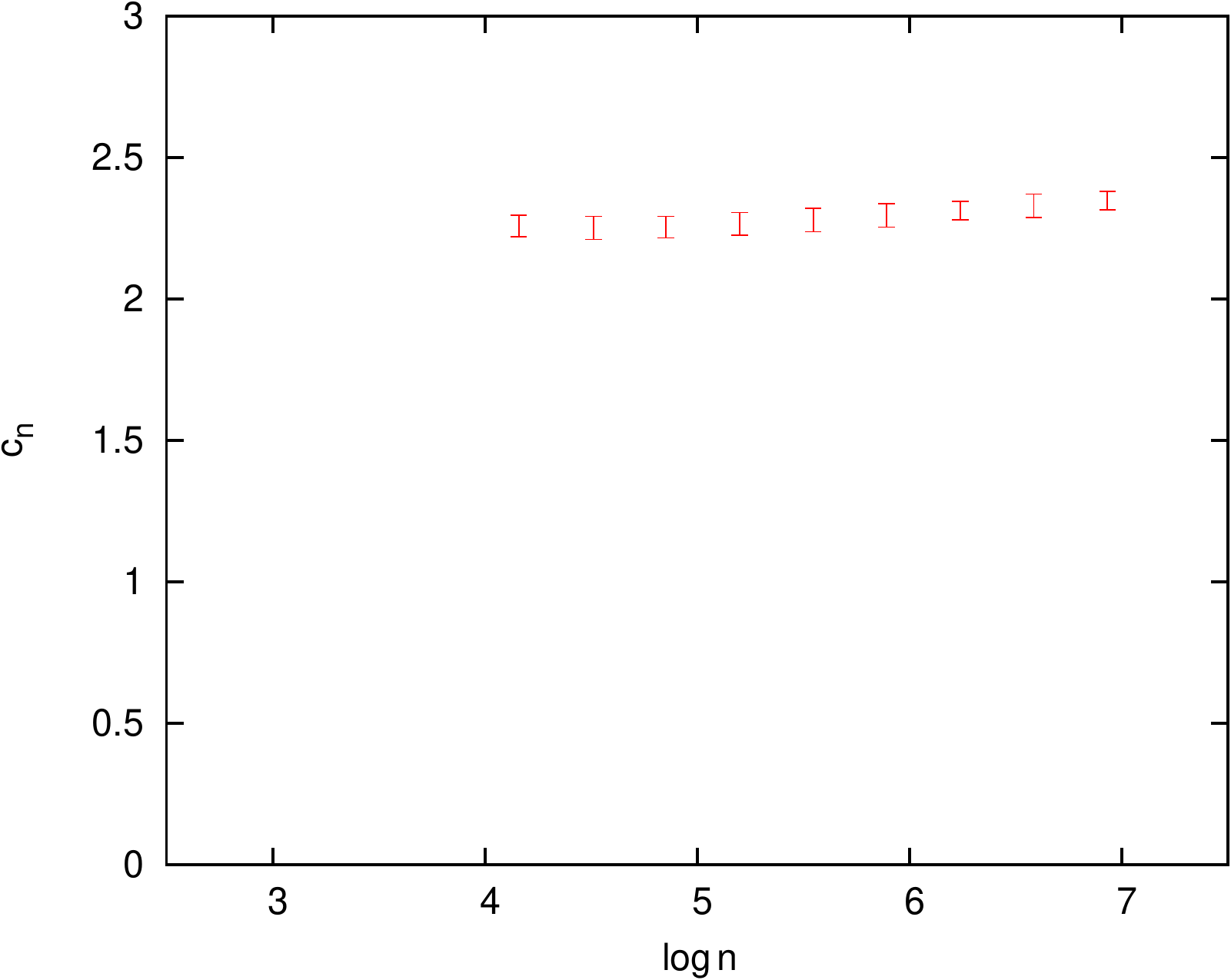}
\end{center}
\caption{Plot of the value of the maximum of the specific heat $c_n= \max_\omega c^{(2)}_n$ against $\log n$. This suggests that the specific heat does not diverge as the polymer length is increased. }
\label{CN_vs_logN} 
\end{figure}

By considering the specific heat we find a weak phase transition in contrast to that found on the square lattice. There is little sign that the specific heat diverges: in Figure~\ref{CN_vs_logN} the value of the maximum of the specific heat is plotted against $\log n$.

\begin{figure}
\begin{center}
\includegraphics[width=0.9\columnwidth]{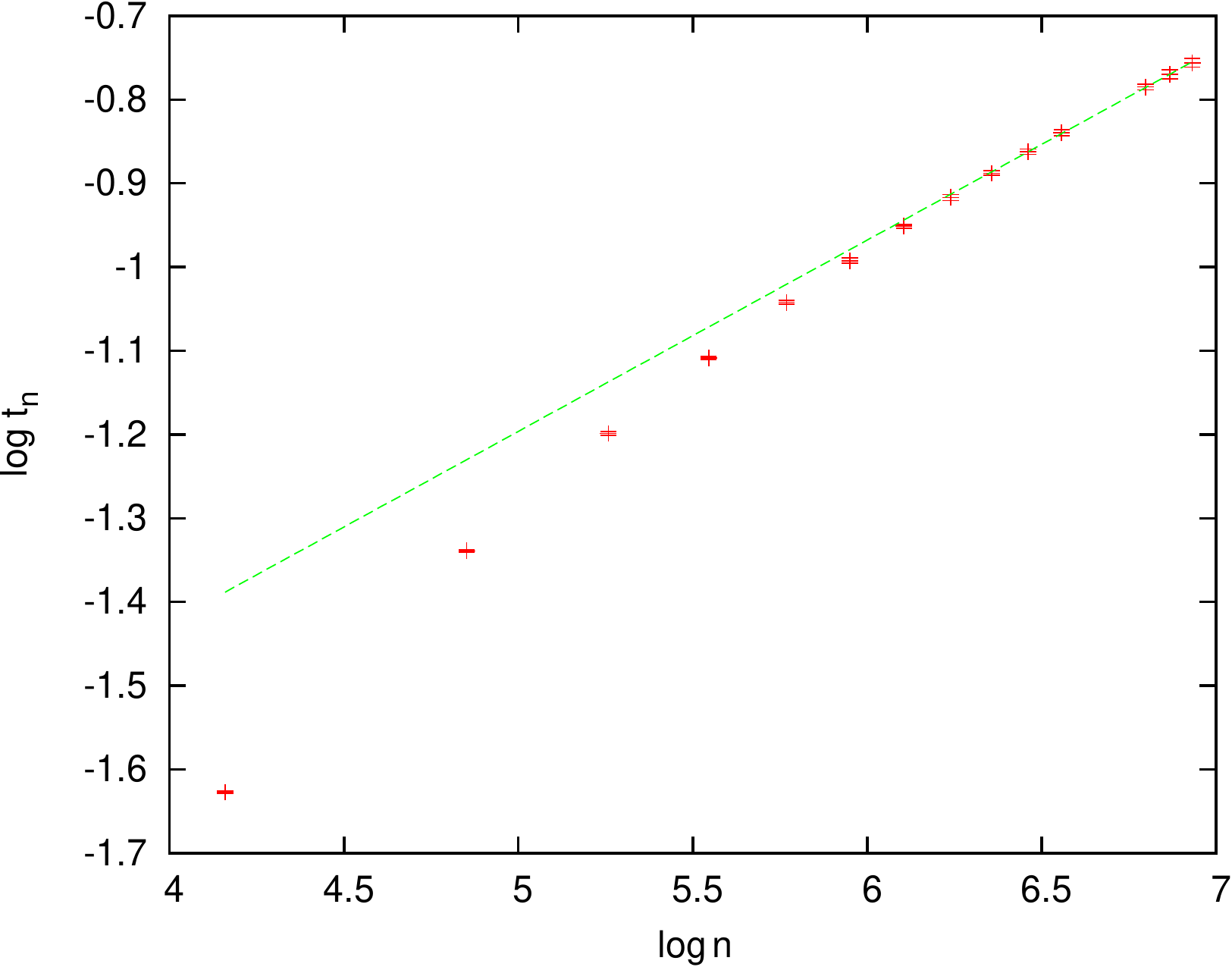}\\
\includegraphics[width=0.9\columnwidth]{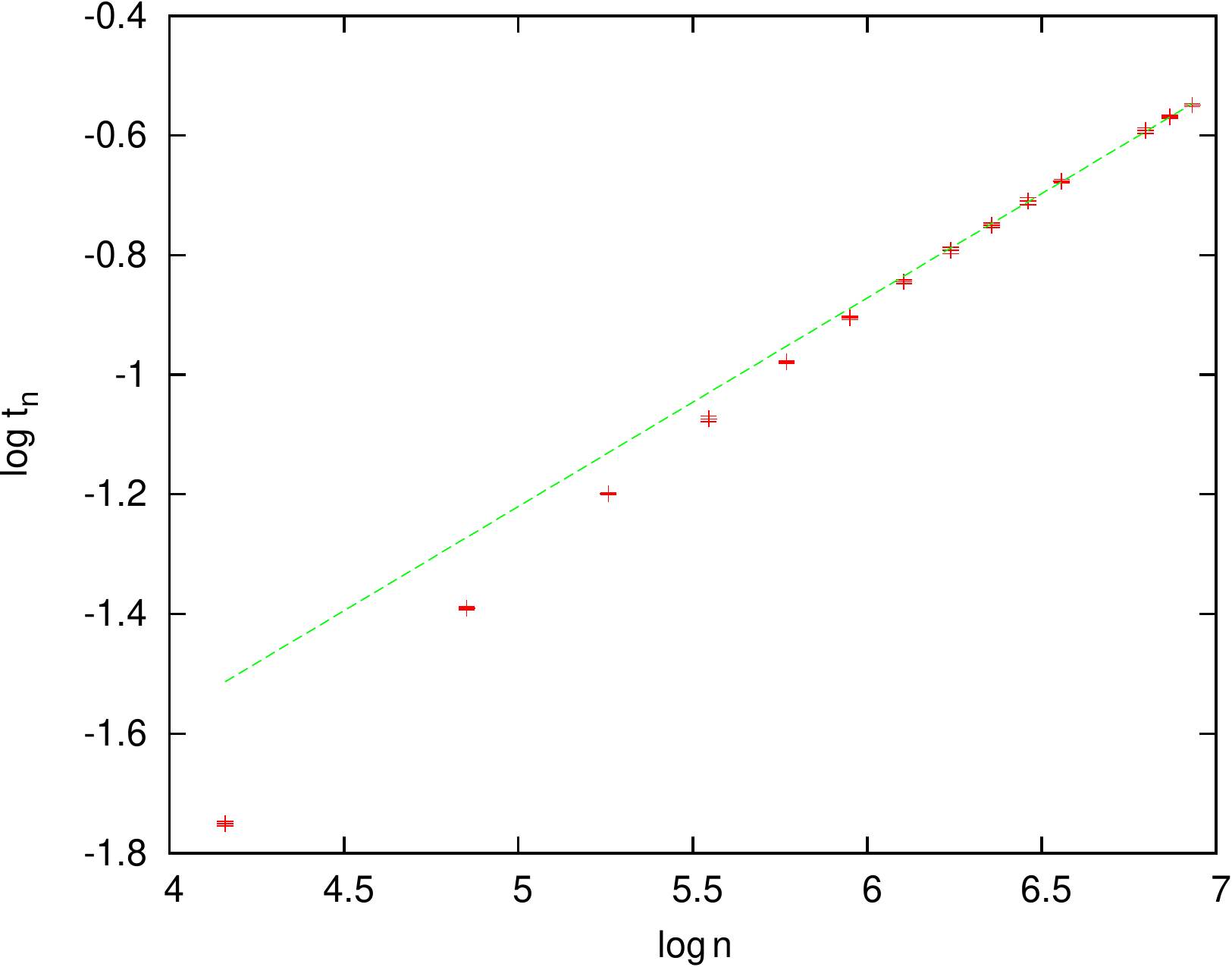}
\end{center}
\caption{Plot of the height of the peaks of $t_n^{(2)}(\omega)$, the third derivative of the free energy with respect to temperature against $n$. The third derivative has two peaks: one positive and one negative in value. The top figure shows $t_n=\max_\omega t_n^{(2)}$ and the bottom figure shows $t_n=\min_\omega t_n^{(2)}$.
}
\label{third_moment_k_2} 
\end{figure}

We recall that the $\theta$-point transition of ISAW has this non-divergent specific heat behaviour, and so we have considered the third derivative of the free energy with respect to $\beta$, $t^{(2)}_n(\omega)$, which would diverge slowly for the $\theta$-point.  The third derivative of the free energy is the first derivative of the specific heat. In Figure~\ref{third_moment_k_2} we display the absolute value of the two peaks  of the third free energy derivative. The third derivative has two peaks: one positive and negative in value. They show a weak divergence: we have extracted local exponents in both cases, which would be given by $(1+\alpha)\phi$, were this a critical point. We find the values $0.23(6)$ and $0.35(6)$: this is consistent with the ISAW $\theta$-point, where $(1+\alpha)\phi=2/7\approx 0.28$. While there are clearly very strong corrections to scaling still in the data, no divergence is found in the specific heat and a weak one in the third derivative. Therefore it is tempting to conjecture that the canonical ISAT model on the triangular lattice has a collapse transition that lies in the $\theta$-point universality class, rather than square lattice ISAT collapse universality class.

\section{Extended ISAT model on the triangular lattice}
\subsection{The model}
The extended model of self-interacting trails (eISAT) on the triangular  lattice is
defined as follows. Consider all different bond-avoiding paths
$\varphi_n$ (open trails and loops) of length $n$ that can be formed on the triangular lattice
with one end fixed at a particular site (the set $\Omega_n$). Associate
an energy $-\varepsilon_2$ with each doubly-visited site and a different energy $-\varepsilon_3$ with each triply-visited site. 
For each configuration $\varphi_n$ count the number $m_2(\varphi_n)$ of
doubly-visited sites and $m_3(\varphi_n)$ of
triply-visited sites  of the lattice and give that configuration a
Boltzmann weight $\omega_2^{m_2}\omega_3^{m_3}$, where $\omega_j=\exp(\beta\varepsilon_j)$. The
partition function of the eISAT model is then given by
\begin{equation}
Z_n(\omega_2,\omega_3) = \sum_{\varphi_n\in \Omega_n} \omega_2^{m_2(\varphi_n)}\omega_3^{m_3(\varphi_n)} \; .
\end{equation}
The probability of a configuration $\varphi_n$ in the equilibrium model is
\begin{equation}
p_{E}(\varphi_n;\omega_2,\omega_3) = \omega_2^{m_2(\varphi_n)}\omega_3^{m_3(\varphi_n)}/Z_n(\omega_2,\omega_3)\; .
\end{equation}
The average of any quantity $Q$ over the
ensemble set of allowed paths $\Omega_n$ of length $n$ is given
generically by
\begin{equation}
\langle Q \rangle(n;\omega_2,\omega_3)= \sum_{\varphi_n\in\Omega_n} Q(\varphi_n)\;  p_{E}(\varphi_n) \; .
\label{extended-isat-average}
\end{equation}
Let us define the expectation values of the number of $j$-fold visited sites per unit length $u_j$ for the equilibrium model as
\begin{equation}
u_j(n;\omega_2,\omega_3) = \frac{\langle m_j\rangle}{n} = \frac{1}{n} \sum_{\varphi(n)} p_E(\varphi(n))\;  m_j(\varphi(n))\;,
\end{equation}
and the fluctuations in these as $c_j$ with
\begin{equation}
c_j (n;\omega_2,\omega_3)  = \frac{\langle m^2_j\rangle  - \langle m_j\rangle^2}{n} \; .
\end{equation}
Note that from Equation~(\ref{m-constraint}), which holds for trail configuration regardless of how they are generated, we have 
\begin{equation}
u_1+2u_2+3u_3 =1\;.
\label{u-constraint}
\end{equation}
The canonical ISAT model is then given by the restriction
\begin{equation}
\omega_2 = \omega \quad \mbox{ and } \quad \omega_3=\omega^2\; .
\end{equation}
By fixing the ratio of the energies $\varepsilon_3/\varepsilon_2 = k$ we have the generalisation 
\begin{equation}
\omega_2 = \omega \quad \mbox{ and } \quad \omega_3=\omega^k\;,
\label{defn-k-model}
\end{equation}
which gives a family parametrised by $k$ of  generalised one-parameter ISAT models. This gives the partition function of our fixed $k$ versions of the eISAT model as 
\begin{equation}
Z_n^{(k)}(\omega) = \sum_{\varphi_n\in \Omega_n} \omega^{m_2(\varphi_n) +k m_3(\varphi_n)} \; .
\end{equation}
Setting $\varepsilon_2=1$ the internal energy is
\begin{equation}
u^{(k)}_n(\omega) =  \frac{\langle (m_2 +k m_3) \rangle}{n}
 \end{equation}
 and the (reduced) specific heat is 
 \begin{equation}
 c^{(k)}_n(\omega)= \frac{\langle (m_2 +k m_3)^2\rangle  - \langle (m_2 +k m_3) \rangle^2}{n} \; .
 \end{equation}
The canonical ISAT model has $k=2$.

We have simulated the general eISAT with a two-parameter FlatPERM algorithm \cite{prellberg2004a-a} up to lengths $128$, generating $8.8.6\times10^7$ samples at that length. 
We have also simulated various specific sub-cases for fixed values of $k$ via one-parameter FlatPERM algorithm typically up to length $1024$, generating in each case roughly $10^7$ samples at that length.

\subsection{Mapping of the growth process to the equilibrium model}
The growth process on the square lattice has been mapped \cite{meirovitch1989d-a,bradley1990a-a} to a specific Boltzmann weight of the equilibrium model  by considering loops. 
The probability distribution $p_G(\varphi_n)$ of the growth process on the triangular lattice can be rewritten in terms of $m_j(\varphi_n)$ as
\begin{align}
p_G (\varphi_n) & = \frac{1}{6} \left( \frac{1}{5}\right)^{m_1(\varphi_n)-1}  \left( \frac{1}{15}\right)^{m_2(\varphi_n)}  \left( \frac{1}{15}\right)^{m_3(\varphi_n)}\nonumber \\
&=  \frac{1}{6} \left( \frac{1}{5}\right)^{n-1}  \left( \frac{5}{3}\right)^{m_2(\varphi_n)}  \left( \frac{25}{3}\right)^{m_3(\varphi_n)} 
\end{align}
using the relationship given in Equation~(\ref{m-constraint}). We then notice that by setting $\omega_2=5/3$ and $\omega_3=25/3$ the equilibrium model has probability distribution
\begin{equation}
p_{E}\left(\varphi_n;\frac{5}{3},\frac{25}{3}\right) = \frac{1}{Z_n(\frac{5}{3},\frac{25}{3})} \; \left(\frac{5}{3}\right)^{m_2(\varphi_n)} \left( \frac{25}{3}\right)^{m_3(\varphi_n)}
\end{equation}
and so we can deduce 
\begin{equation}
p_G (\varphi_n)\propto p_{E}\left(\varphi_n;\frac{5}{3},\frac{25}{3}\right)\; .
\end{equation}
Note that the normalisation is different though since the sum over all walks of fixed length $n$ gives the probability of walks being still open in the case of the growth process, and unity in the case of the equilibrium model. 

We now consider the probability of producing configurations in the growth process conditioned on the process having continued to length $n$, that is, we consider
\begin{equation}
\hat{p}_G(\varphi_n) = \frac{p_G(\varphi_n)}{P_n}
\end{equation}
which implies that
\begin{equation}
\sum_{\varphi_n \in \Omega_n}\hat{p}_G(\varphi_n) = 1\; .
\end{equation}
Hence we have
\begin{equation}
\hat{p}_G (\varphi_n)= p_{E}\left(\varphi_n;\frac{5}{3},\frac{25}{3}\right)\;,
\end{equation}
noting that this implies that 
\begin{equation}
Z_n\left(\frac{5}{3},\frac{25}{3}\right) = 6 \,5^{n-1} P_n\; .
\end{equation}
In any case, when we simulate the growth process we are effectively simulating the equilibrium eISAT model at the point 
$(\omega_2,\omega_3)=(5/3,25/3)$. We immediately have that
\begin{equation}
e_j(n) = u_j\left(n;\frac{5}{3},\frac{25}{3}\right)  \quad \mbox{ and } \quad f_j(n) = c_j\left(n;\frac{5}{3},\frac{25}{3}\right)\; .
\end{equation}
Using the definition of the one-parameter family of models via Equation~(\ref{defn-k-model}), the growth process is equivalent to $\omega=5/3$ in the model where 
\begin{equation}
k=k_G\equiv\frac{\log(25/3)}{\log(5/3)}\approx 4.15\;.
\end{equation}
Importantly, this is not the value of $k$ for the canonical ISAT model (with $k=2$) on the triangular lattice that has been studied via exact enumeration \cite{lim1988a-a}.

\subsection{The $k=k_{G}$ eISAT model}
\begin{figure}[ht!]
\begin{center}
\includegraphics[width=0.9\columnwidth]{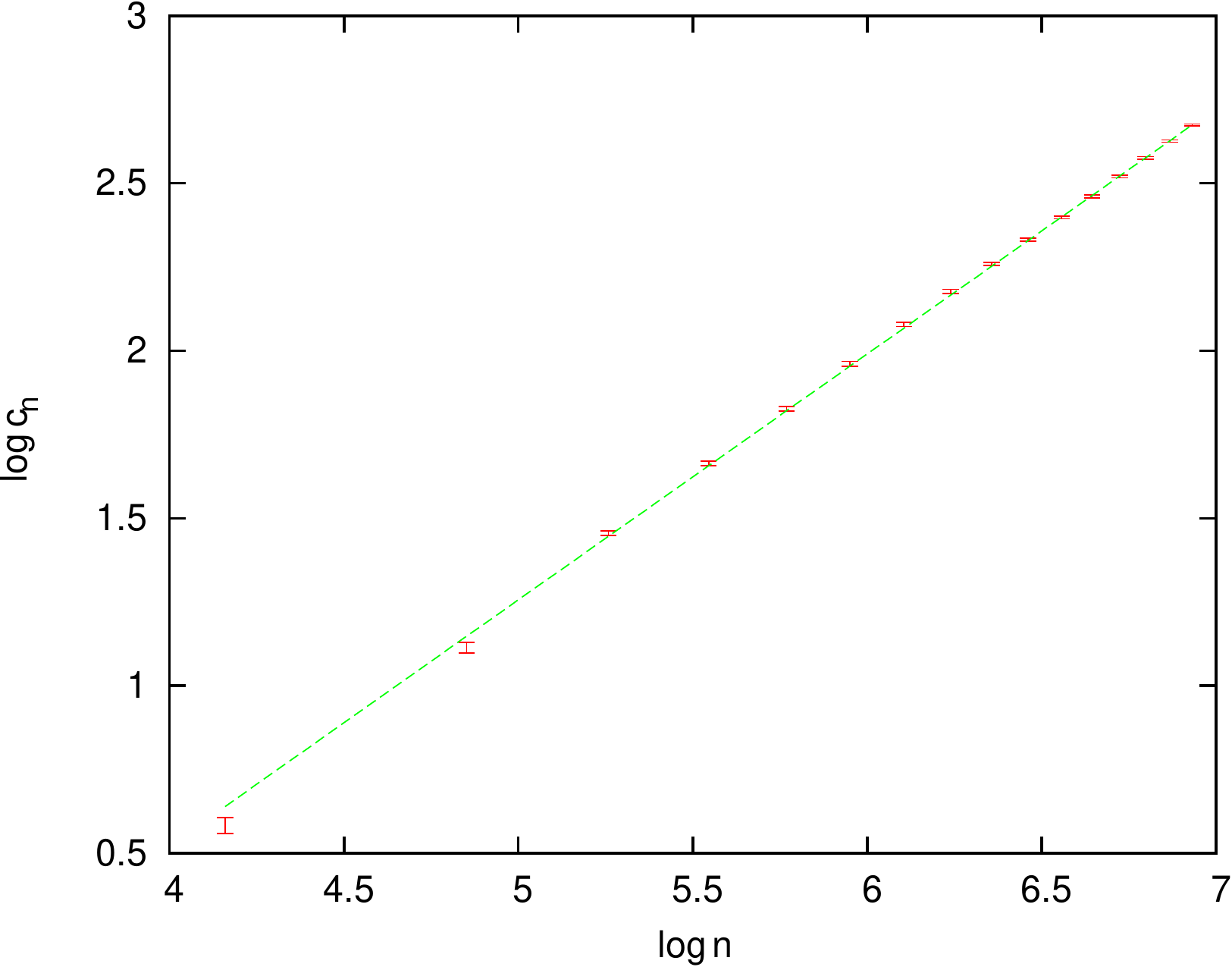}
\end{center}
\caption{Plot of the logarithm of $c_n= \max_\omega c^{(k_G)}_n$, the value of the maximum of the specific heat, against $\log n$. The straight line has slope $0.734$.}
\label{specific_heat_k_kgt} 
\end{figure}

\begin{figure}[ht!]
\begin{center}
\includegraphics[width=0.9\columnwidth]{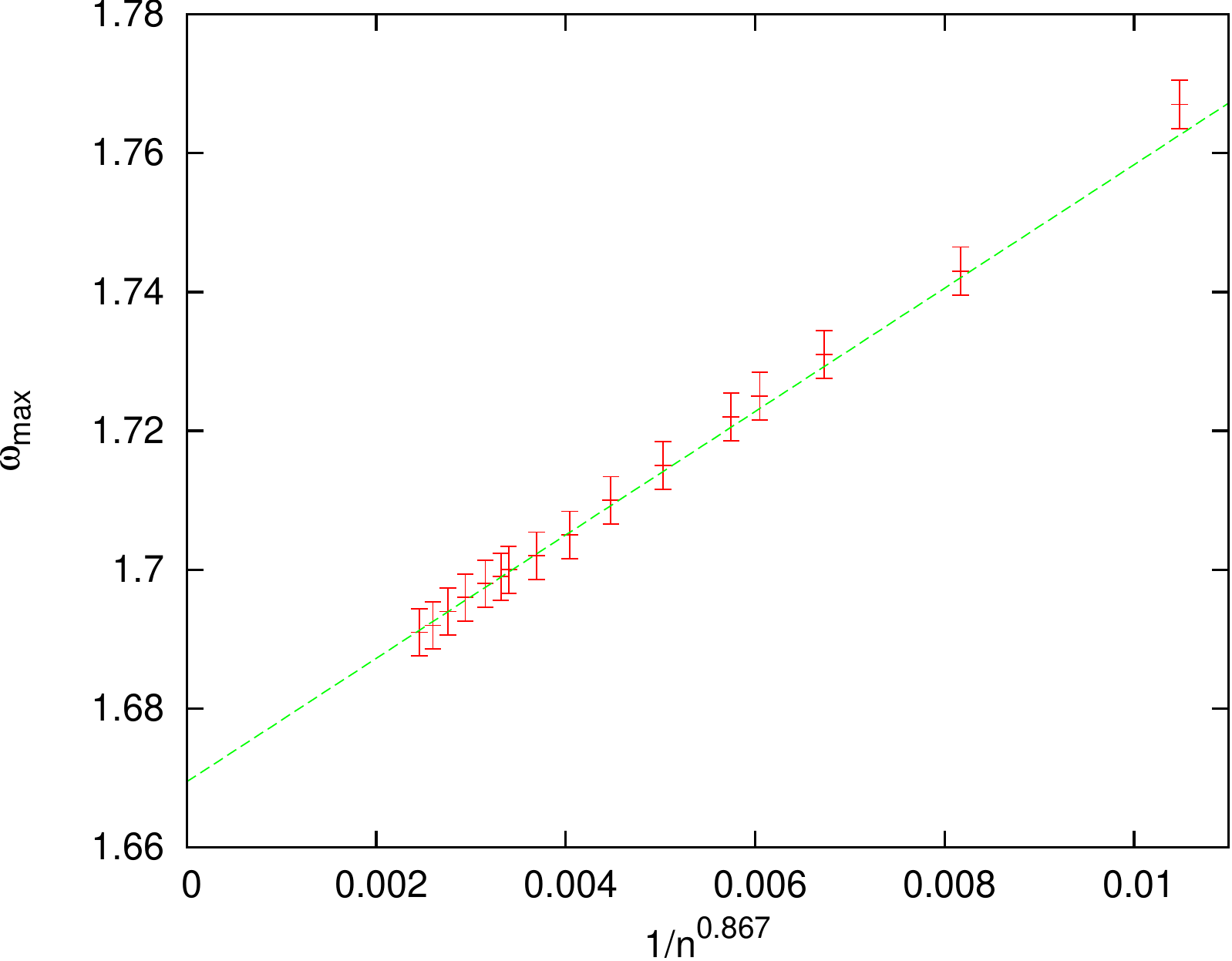}
\end{center}
\caption{Plot of the location, $\omega_{max}$, of the peak of the specific heat against $1/n^{0.867}$.}
\label{omegainf_k_kgt}
\end{figure}

As described in Section 2 we have simulated the growth process and found divergent fluctuations in the number of doubly and triply-visited  sites --- a sign of critical behaviour. 
Now that we have mapped the growth process onto a specific temperature of the $k=k_G$ eISAT model we can verify that this point is indeed the collapse transition point. We have simulated the $k=k_G$ model up to length $1024$ using a one-parameter FlatPERM algorithm. We find a divergent specific heat with a single pronounced peak near $\omega=5/3$. We begin by finding the location and size of the peak of the specific heat of the model. In Figure~\ref{specific_heat_k_kgt} we plot the logarithm of the peak height of the specific heat against $\log n$ along with a line corresponding to the exponent value $\alpha\phi =0.734$, which was obtained from the growth process. Moreover, using the finite length estimate of $\phi=0.867$ obtained from the growth process to extrapolate the location of the transition (see Figure~\ref{omegainf_k_kgt}) we find an estimate of $\omega_t=1.669(4)$: this is consistent with the growth process point at $\omega=1.6\dot{6}$ being the transition point.

\begin{figure}[ht!]
\begin{center}
\includegraphics[width=0.9\columnwidth]{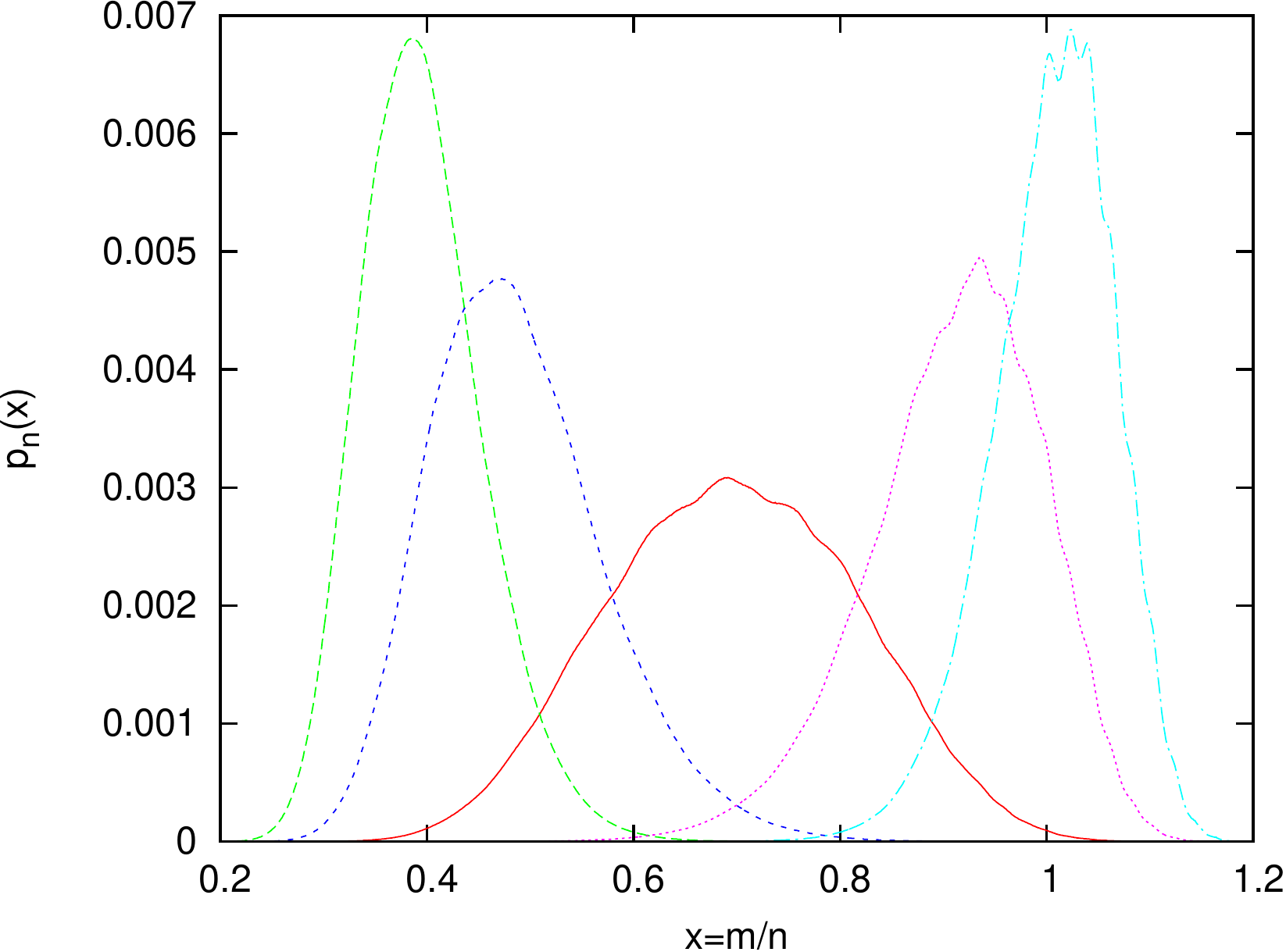}
\end{center}
\caption{Plot of the distribution $p_n(m/n)$, where $m=m_2+k_Gm_3$, at temperatures near, and at, the temperature at which the specific heat attains its maximum for length $n=1024$. The specific heat attains its maximum at $\omega=\omega_{max}=1.69$ and the distribution is plotted for this value and at $\omega = 1.63,1.66,1.72,1.76$: the plots move from left to right as $\omega$ is increased.}
\label{prob_1024_k_kgt} 
\end{figure}

While the exponent estimates clearly discount a first-order transition, given that there was a question about the first-order nature of the ISAT model  on the square lattice at the kinetic growth point we now show how the distribution of the numbers $m=m_2+k_Gm_3$ changes as the temperature is moved through the transition point: this is displayed in Figure~\ref{prob_1024_k_kgt}.
There is no sign of first-order behaviour in the distributions.

\begin{figure}[ht!]
\begin{center}
\includegraphics[width=0.9\columnwidth]{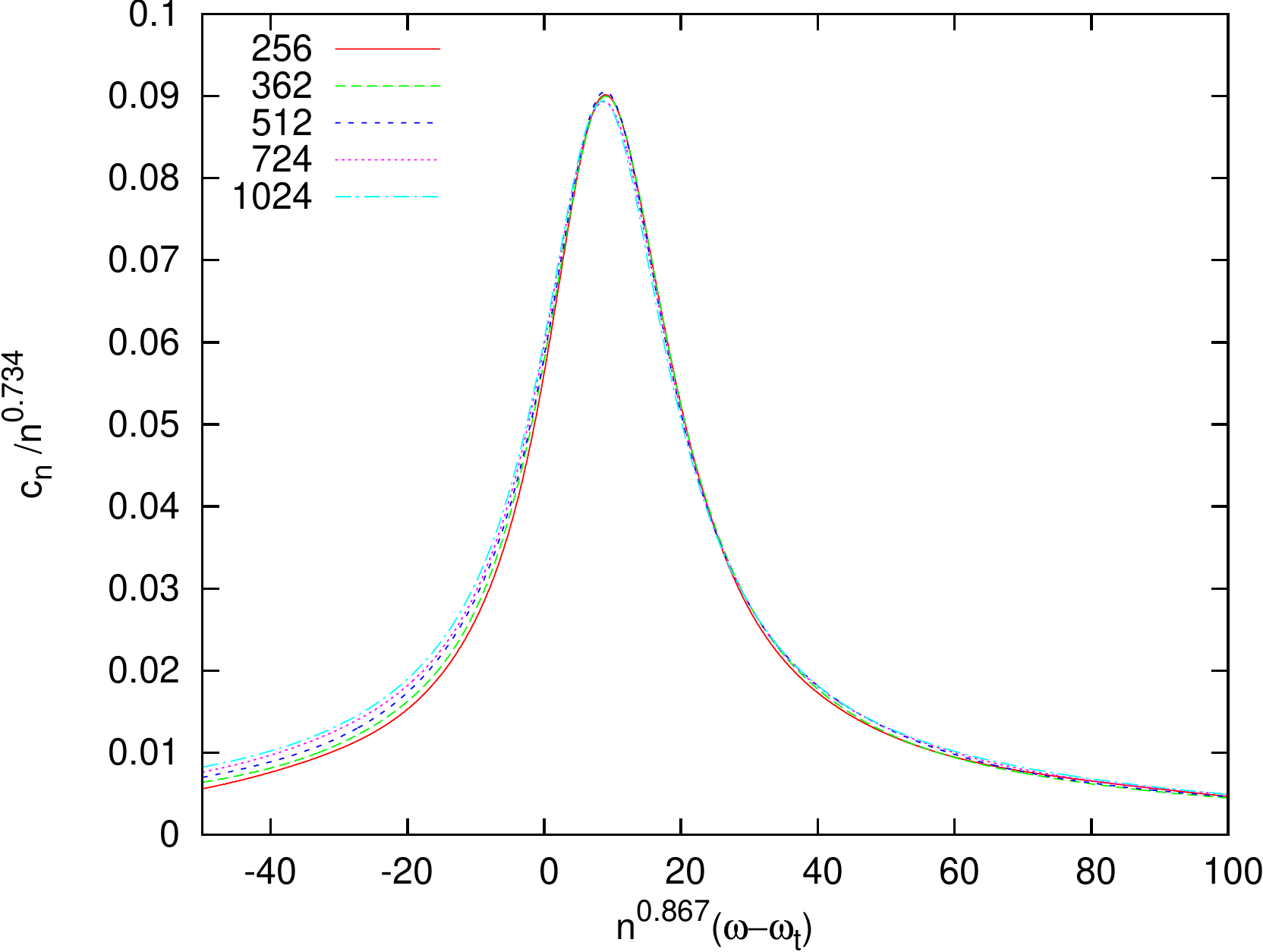}
\end{center}
\caption{Scaling plot of the specific heat around the transition temperature, using the exponents from the growth process.}
\label{scaled_C} 
\end{figure}

To complete our numerical analysis we provide a scaling plot around the transition of the specific heat using the exponents values from the growth process. This is found in Figure~\ref{scaled_C} and it shows an excellent fit to the crossover scaling form \cite{brak1993a-:a}
\begin{equation}
c_n \sim n^{\alpha\phi} {\cal C}((\omega - \omega_t)n^\phi)\;.
\end{equation}
Our data at length $1024$ there fully confirms that the kinetic growth point is the location of a phase transition and, moreover, that this transition is critical in nature. The exponents values indicate that the ISAT model on the square lattice and the $k=k_G$ eISAT model on the triangular lattice are in the same universality class. This is in contrast to the observation above that the canonical ISAT models on the two lattices are \emph{not} in the same universality class.
This inexorably leads us to the conclusion that the universality class of the phase transition of the eISAT model on the triangular lattice depends on the value of $k$.

\subsection{The $k=0$ eISAT model}
\begin{figure}[ht!]
\begin{center}
\includegraphics[width=0.9\columnwidth]{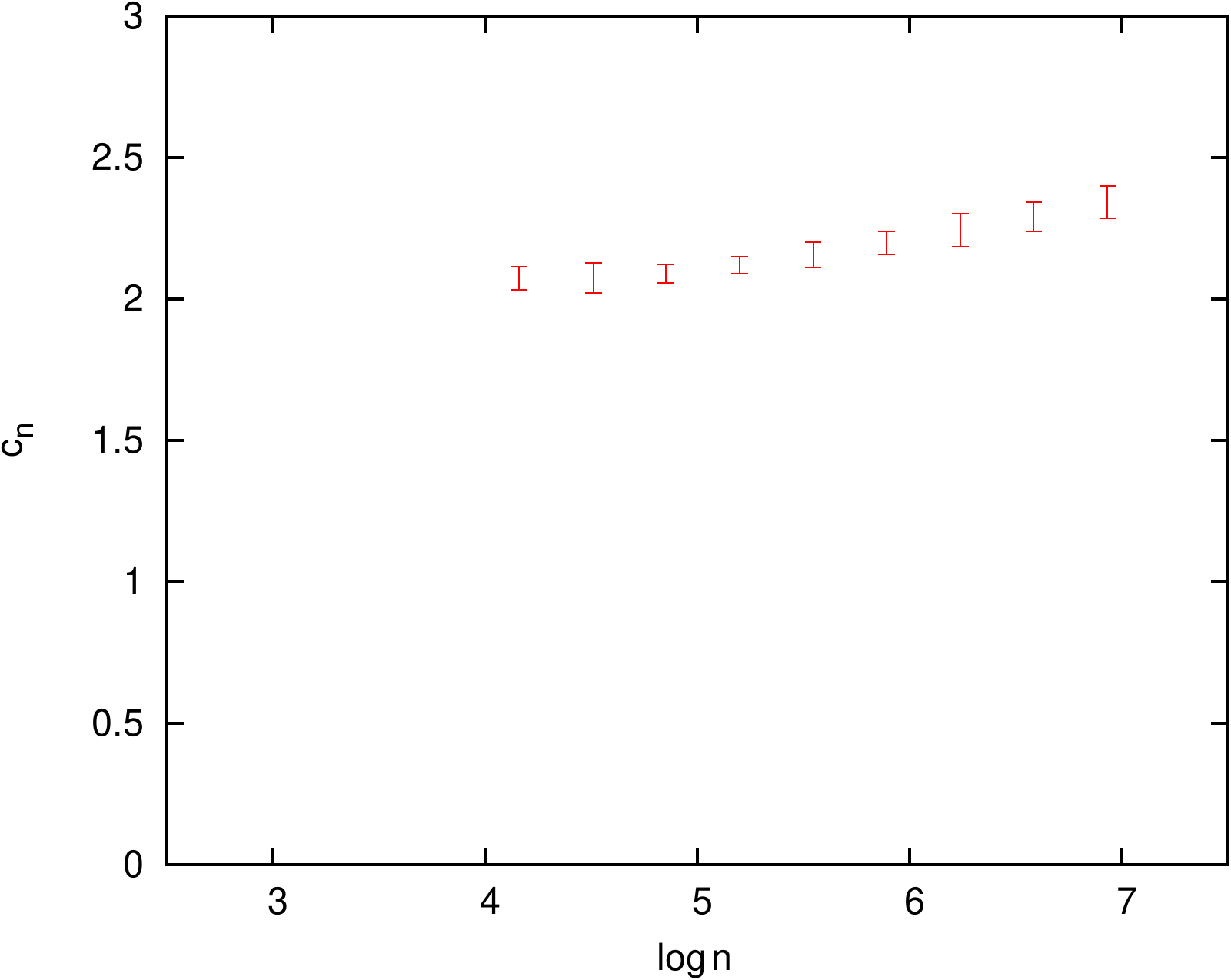}
\end{center}
\caption{Plot of the value of the maximum of the specific heat $c_n= \max_\omega c^{(0)}_n$ against $\log n$. This suggests that the specific heat does not diverge as the polymer length is increased, as is the case in the canonical model ($k=2$).
}
\label{specific-heat-k-0} 
\end{figure}

\begin{figure}
\begin{center}
\includegraphics[width=0.9\columnwidth]{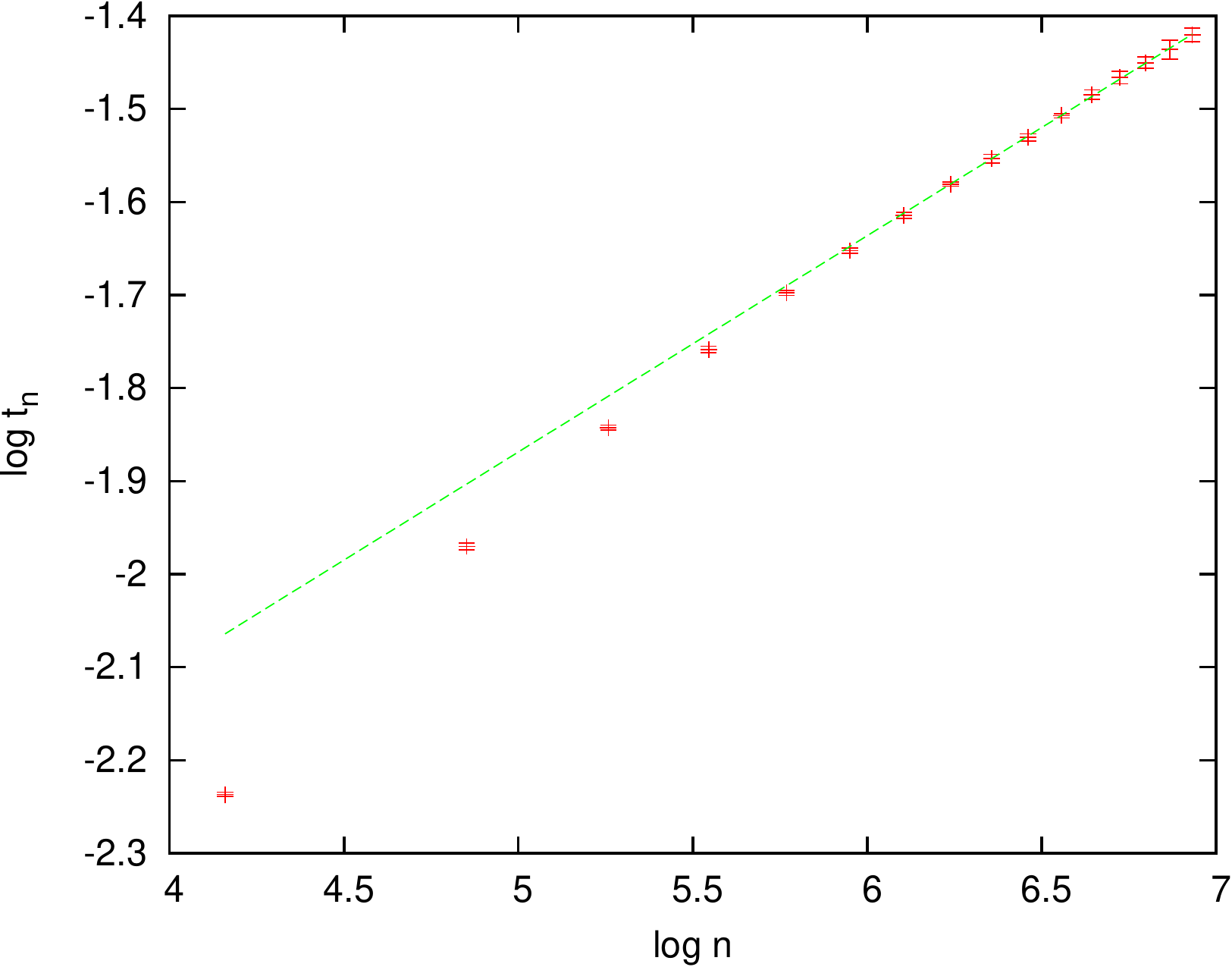}
\end{center}
\caption{
Plot of the height of one of  the peaks of $t_n^{(0)}(\omega)$, the third derivative of the free energy with respect to temperature against $n$. The third derivative has two peaks: one positive and one negative in value. The  figure shows $t_n=\min_\omega t_n^{(0)}$ which is the larger in absolute value and the one seemingly less affected by corrections to scaling.}
\label{third_moment_k_0} 
\end{figure}

To explore this $k$-dependency more we now examine the $k=0$ eISAT model.
When $k=0$ only doubly-visited sites are given a Boltzmann weight so that 
\begin{equation}
Z_n^{(0)}(\omega) = \sum_{\varphi_n\in \Omega_n} \omega^{m_2(\varphi_n)}\; .
\end{equation}
The internal energy is
\begin{equation}
u^{(0)}_n(\omega) =  \frac{\langle m_2 \rangle}{n}
 \end{equation}
 and the specific heat is 
 \begin{equation}
 c^{(0)}_n(\omega)= \frac{\langle m_2^2\rangle  - \langle m_2  \rangle^2}{n} \;.
 \end{equation}
In Figure~\ref{specific-heat-k-0} the value of the maximum  of the specific heat over all $\omega$ is plotted against $\log n$. In stark contrast to the behaviour of the $k=k_G$ model but in common with the $k=2$ canonical model the specific heat does not seem to diverge as the length is increased.
In common with the approach we took for the $k=2$ model, we have also examined the third derivative of the free energy. In Figure~\ref{third_moment_k_0} we see that the absolute maximum of this quantity is  weakly divergent  with an exponent of $0.23(6)$. 
This compares well to the $\theta$-point value of $2/7\approx 0.28$, especially given the relatively short lengths of the simulations. It is tempting then to conjecture that $k=0$ and $k=2$ models have collapse transitions in the same universality class and that this class is the $\theta$-point.

\subsection{The Triple model (``$k=\infty$'')}

Next we consider a model where only triply-visited sites are weighted --- essentially a ``$k=\infty$'' eISAT --- that is, the partition function is given by 
\begin{equation}
Z_n^{(triple)}(\omega) = \sum_{\varphi_n\in \Omega_n} \omega^{m_3(\varphi_n)}\; .
\end{equation}
The internal energy is
\begin{equation}
u^{(triple)}_n(\omega) =  \frac{\langle m_3 \rangle}{n}
 \end{equation}
 and the specific heat is 
 \begin{equation}
 c^{(triple)}_n(\omega)= \frac{\langle m_3^2\rangle  - \langle m_3  \rangle^2}{n} \;.
 \end{equation}
Note that only in the Triple model $\omega=\omega_3$, while for the finite $k$ models we denote $\omega=\omega_2$.

\begin{figure}[ht!]
\begin{center}
\includegraphics[width=0.9\columnwidth]{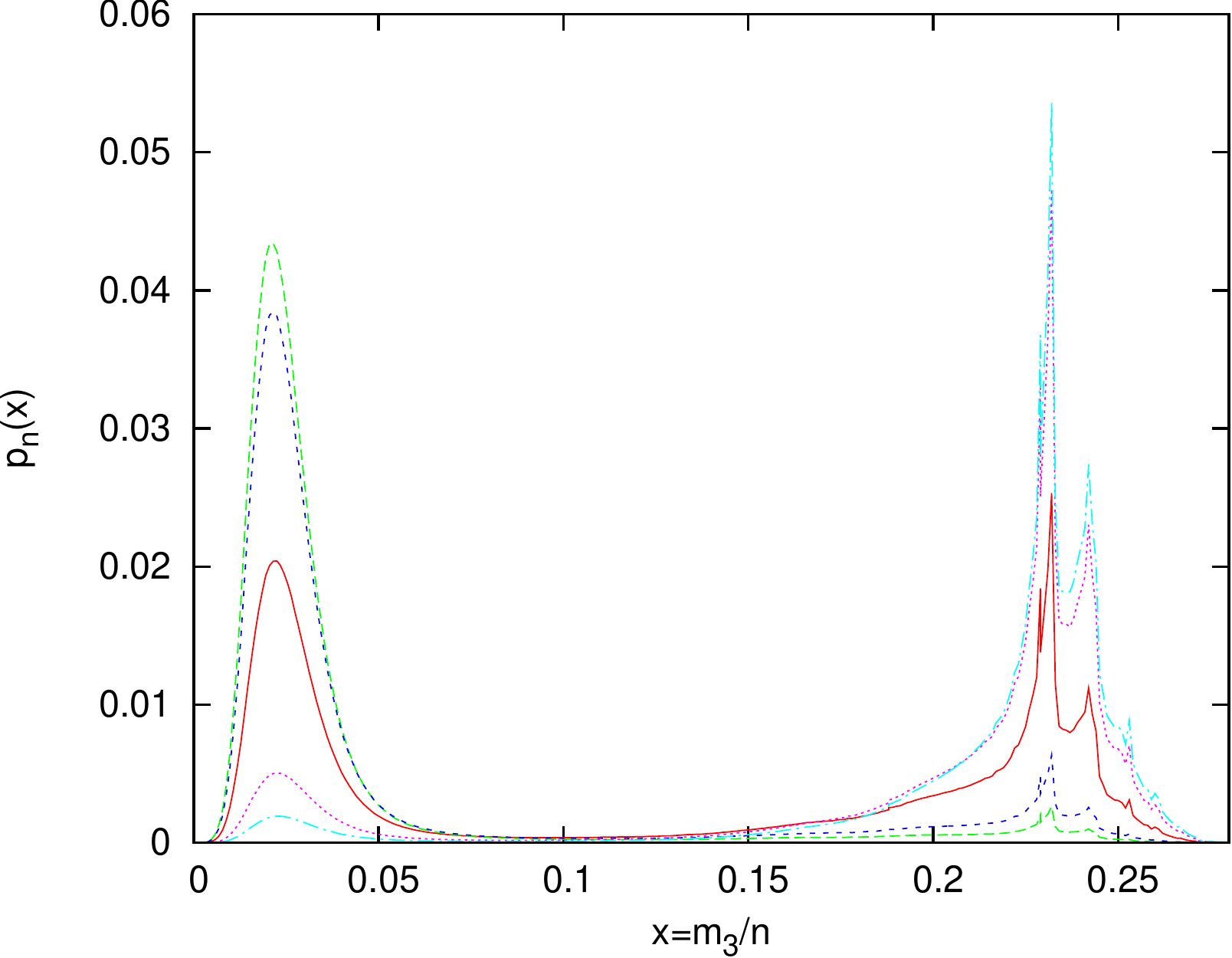}
\end{center}
\caption{Plot of the distribution $p_n(m_3/n)$  of triply-visited sites for the \emph{Triple} model at temperatures near, and at, the temperature at which the specific heat attains its maximum for length $n=1024$. The specific heat attains its maximum at $\omega=\omega_{max}=7.41$ and the distribution is plotted for this value and at $\omega =7.31,7.34,7.48,7.52$: the plots move from left to right as $\omega$ is increased.
}
\label{prob_1024_k_inf} 
\end{figure}

\begin{figure}[ht!]
\begin{center}
\includegraphics[width=0.9\columnwidth]{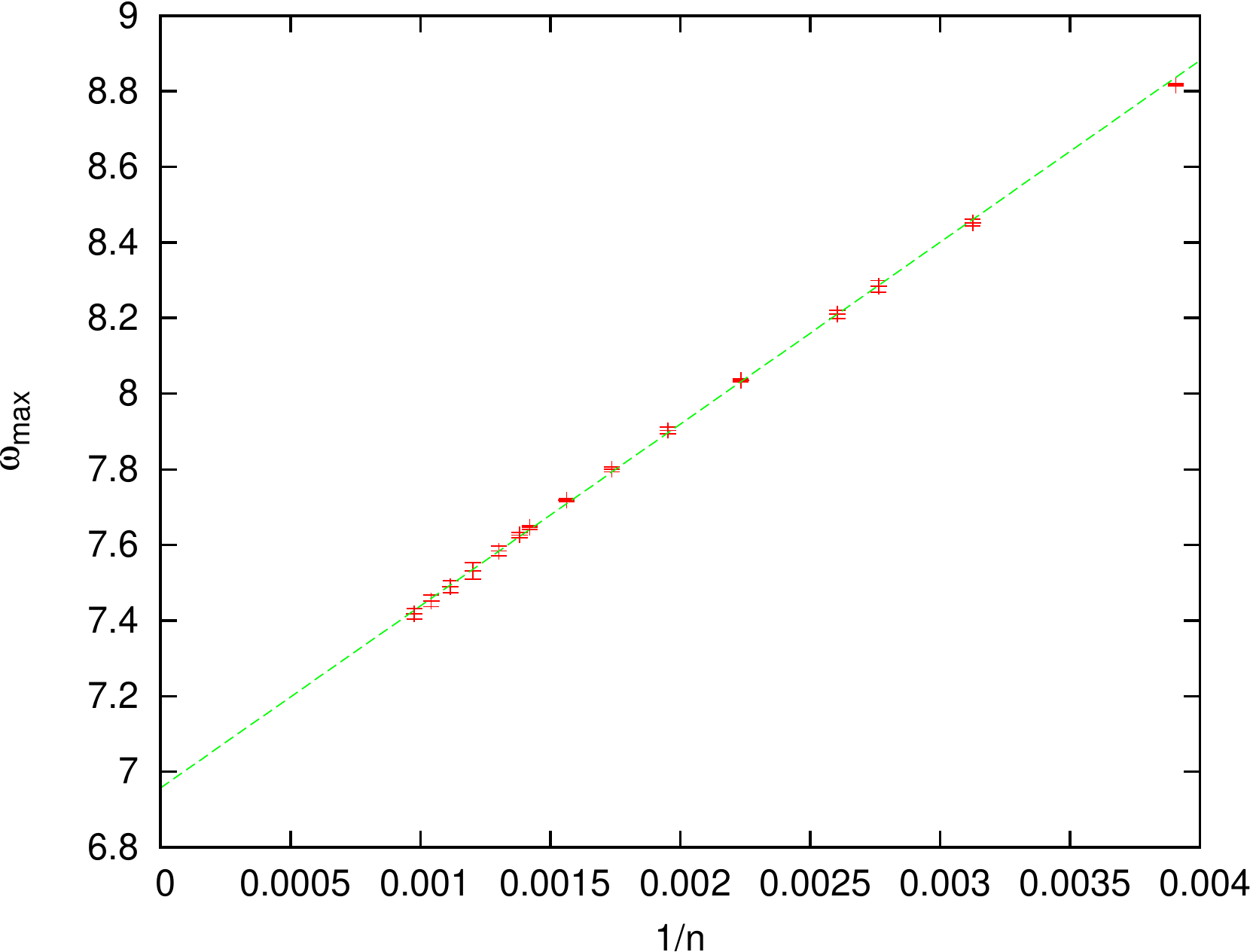}
\end{center}
\caption{Plot of the location, $\omega_{max}$,  of the peak of the specific heat against $1/n$ for the \emph{Triple} model.}
\label{omegainf_k_inf}
\end{figure}

As a function of $\omega$ we find evidence for  a very strong phase transition. In fact when we considered the scaling of the peaks of the specific heat they scale faster than linearly (linearly is the maximum theoretical asymptotic behaviour). Since linear behaviour would indicate a first-order phase transition we considered the distribution of triply-visited sites at temperatures around that which gives rise to the peak of the specific heat: this is plotted in Figure~\ref{prob_1024_k_inf}.
The distribution displays an unambiguous double peaked form which is a clear sign of a first-order phase transition. Note the sharpness of the transition in temperature and how the distribution moves from being peaked around low values of triply-visited sites just below the transition temperature to peaked around large values just above the transition temperature. This is classic first-order behaviour.
To get an idea of where the transition takes place in the thermodynamic limit we have extrapolated the transition location in Figure~\ref{omegainf_k_inf}.
We estimate a thermodynamic limit value of $\omega_t(\text{\emph{Triple}})= 6.96(6)$. 

We now see that when triply-visited sites are weighted the transition is clearly first-order, while when doubly-visited sites are the only ones given additional weight the transition is a weak second-order transition, not unlike the $\theta$-point. In between, the growth model displays intermediate behaviour. 

\subsection{General eISAT model}

\begin{figure}[ht!]
\begin{center}
\includegraphics[width=0.9\columnwidth]{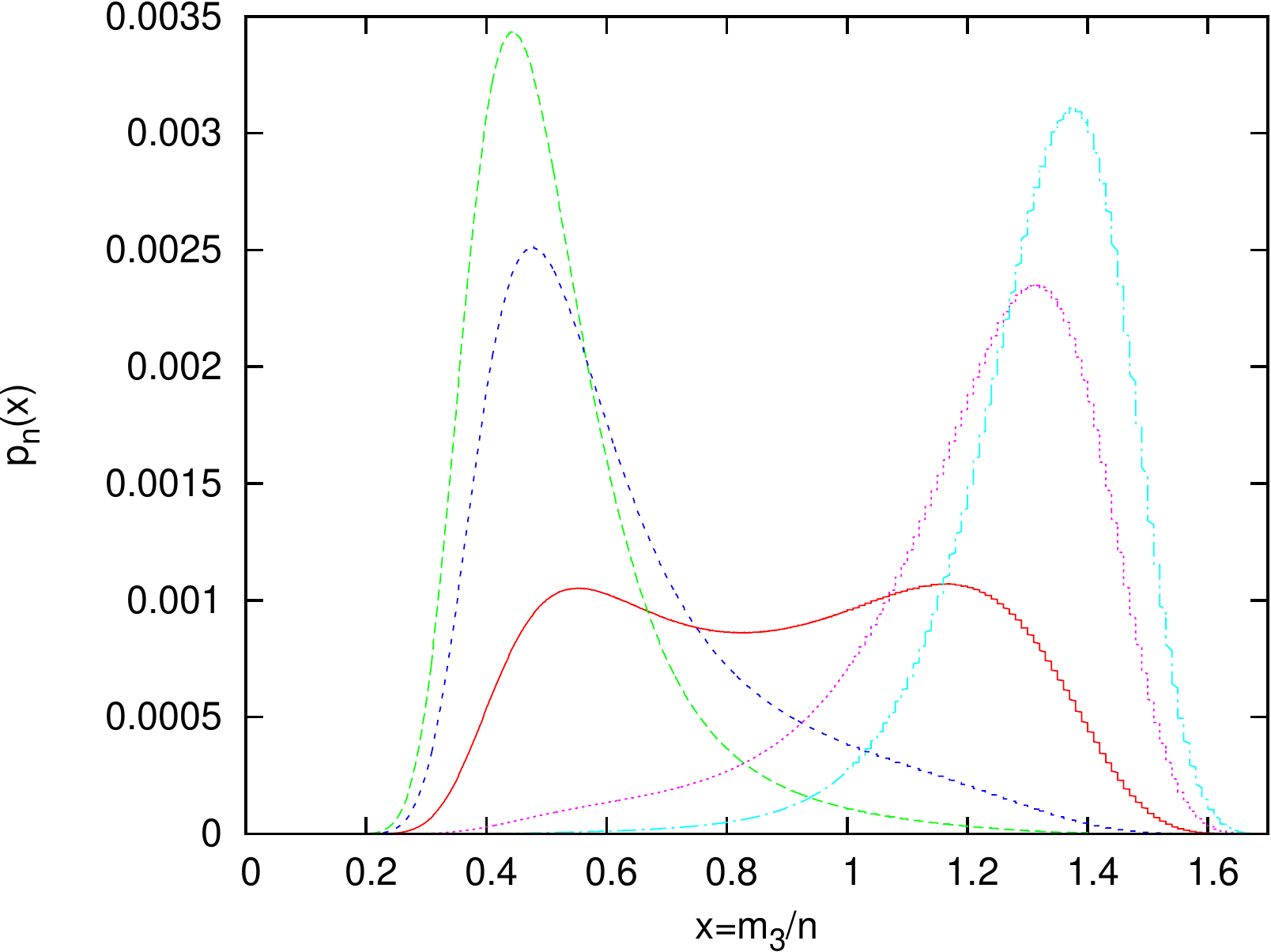}
\end{center}
\caption{Plot of the distribution $p_n(m_3/n)$  of triply-visited sites for the $k=6$ model at temperatures near, and at, the temperature at which the specific heat attains its maximum for length $n=1024$. The specific heat attains its maximum at $\omega=\omega_{max}=1.419$ and the distribution is plotted for this value and at $\omega =1.410,1.414,1.425,1.430$: the plots move from left to right as $\omega$ is increased.}
\label{prob_1024_k_6} 
\end{figure}

To understand whether the first-order nature of the collapse transition persists when doubly-weighted sites are given some weight we have examined the $k=6$ model. Once again the divergence of the specific heat is very strong, and plotting the distribution of triply-visited (see Figure~\ref{prob_1024_k_6}) the classic doubly peaked form is once again apparent.
It seems that the first-order nature persists for large finite values of $k$. Given the evidence available, the simplest scenario that presents itself is the following:  For values of $k>k_G$ the eISAT model on the triangular lattice displays a first-order phase transition, while for $k<k_G$ the model displays a weak second-order transition presumably in the universality class of the ISAW $\theta$-point. Separating these two behaviours is the $k=k_G$ model, where the transition is second-order but with a divergent specific heat in the universality class of square lattice ISAT. This implies that the growth process point is a multi-critical point that is the meeting of a line of first-order transitions to a line of critical ones. 

\begin{figure}[ht!]
\begin{center}
\includegraphics[width=0.9\columnwidth]{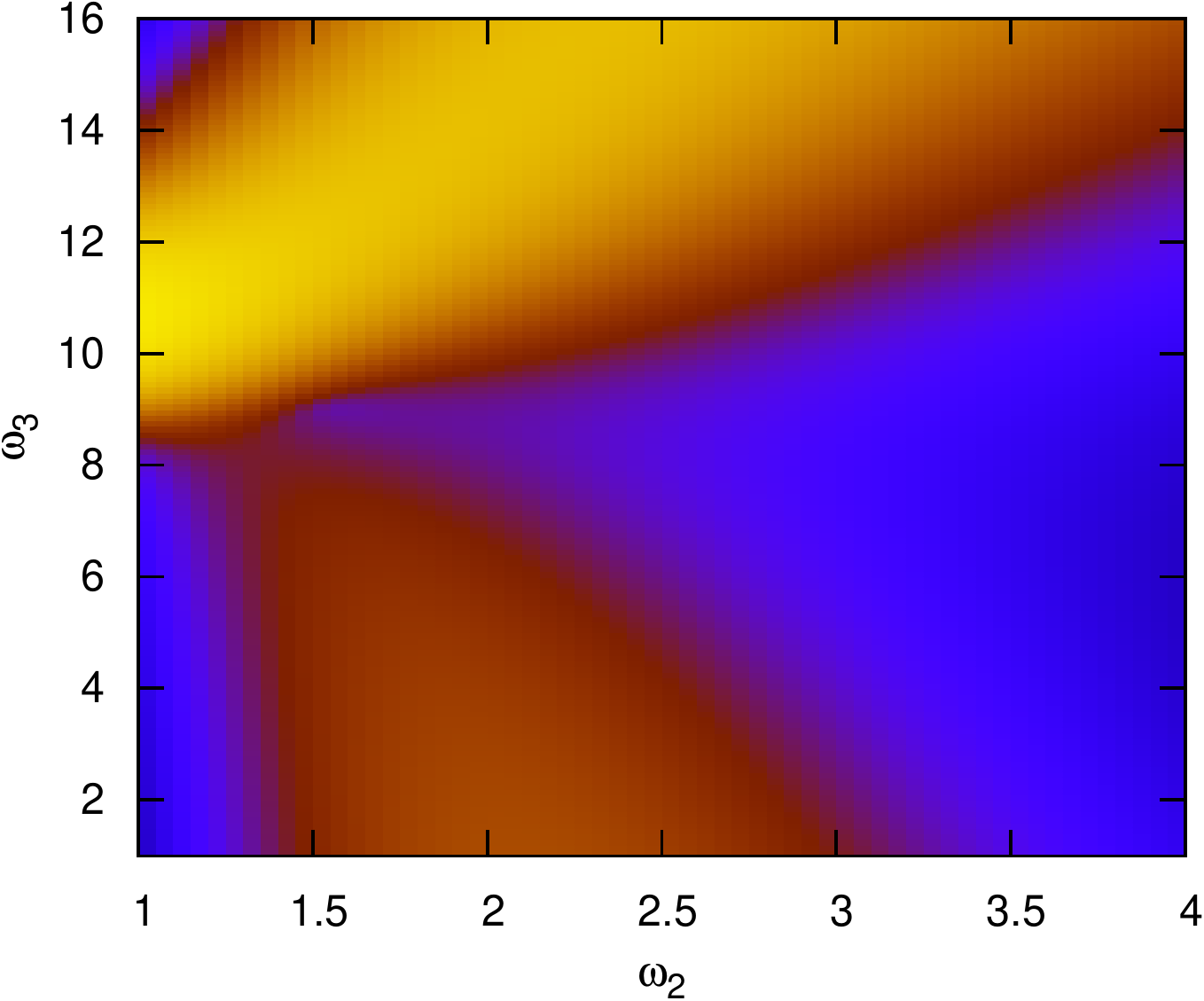}
\end{center}
\caption{Density plot of the logarithm of the largest eigenvalue $\lambda_{max}$ of the matrix
of second derivatives of the free energy with respect to $\omega_2$ and $\omega_3$ at length $n=128$ (the lighter
the shade, the larger the value).}
\label{param_pm3d} 
\end{figure}

The question then arises as to the nature of the low temperature phase for different $k$. In Figure~\ref{param_pm3d} we give a density plot of the largest eigenvalue of the matrix of second derivative of the free energy with respect to the two variable $\omega_2$ and $\omega_3$ at length $n=128$: this allows us to search for any other possible transitions. Intriguingly this plot show evidence for transitions at large $\omega_2$ and $\omega_3$.
 This would seem to indicate a collapse-collapse phase transition with two different collapsed states. Such a transition has been seen  to occur in the low temperature   behaviour of the ISAW when a stiffness energy is added to model semi-flexible polymers. One of the two low temperature phases in that model is an ordered crystalline-like  phase. Relatedly this occurs when the collapse is a first-order transition.

\begin{figure}[ht!]
\begin{center}
\includegraphics[width=0.9\columnwidth]{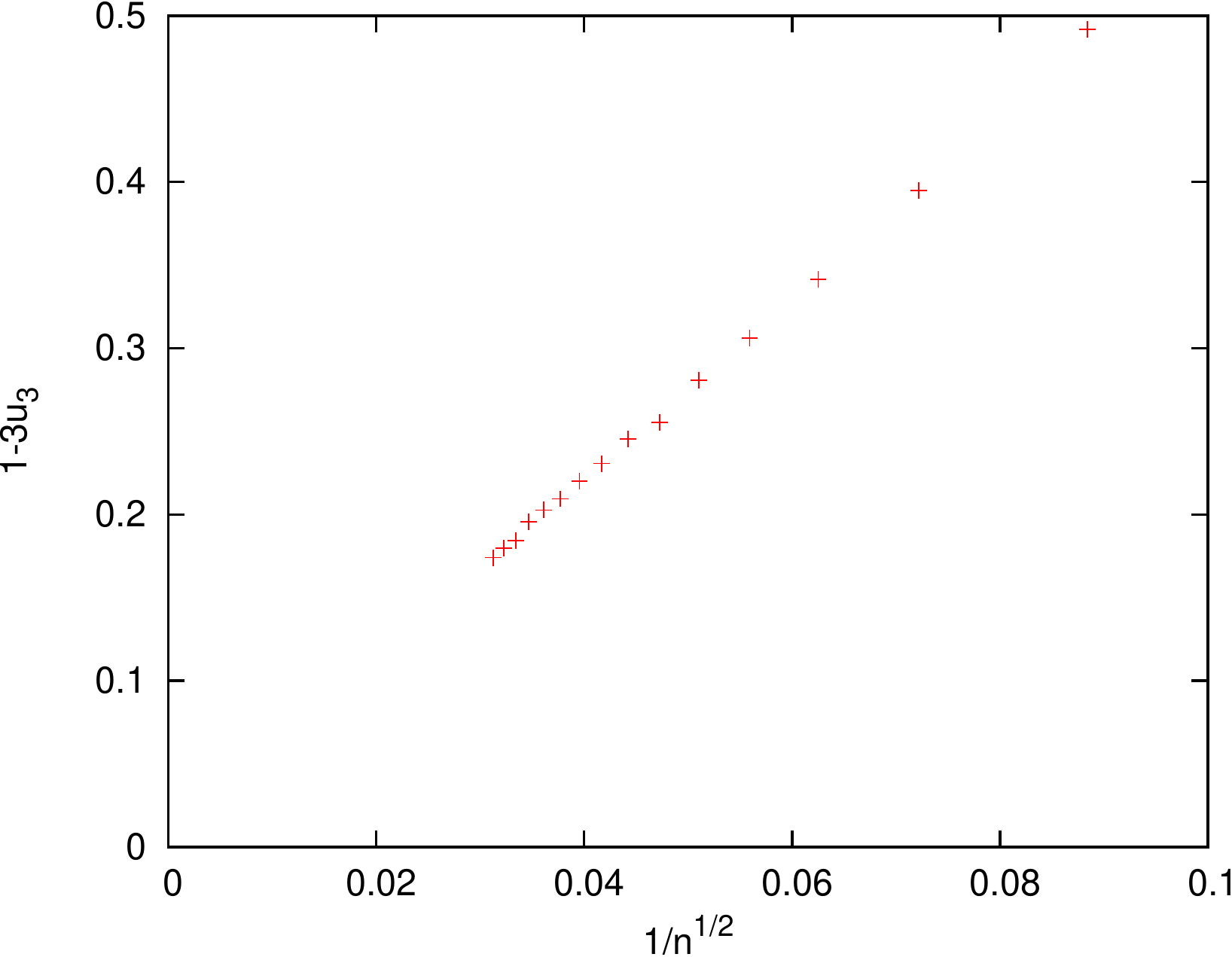}
\end{center}
\caption{Plot of $1-3 u_3(n)$, which measures the proportion of steps that  are not involved with triply-visited sites per unit length, against $1/\sqrt{n}$ at a point $(1.58,15.6)$ in the hypothesised frozen (crystal-like) phase. As the length increases this quantity  vanishes. 
}
\label{plot_density_k_6} 
\end{figure}

\begin{figure}
\begin{center}
\includegraphics[width=0.7\columnwidth]{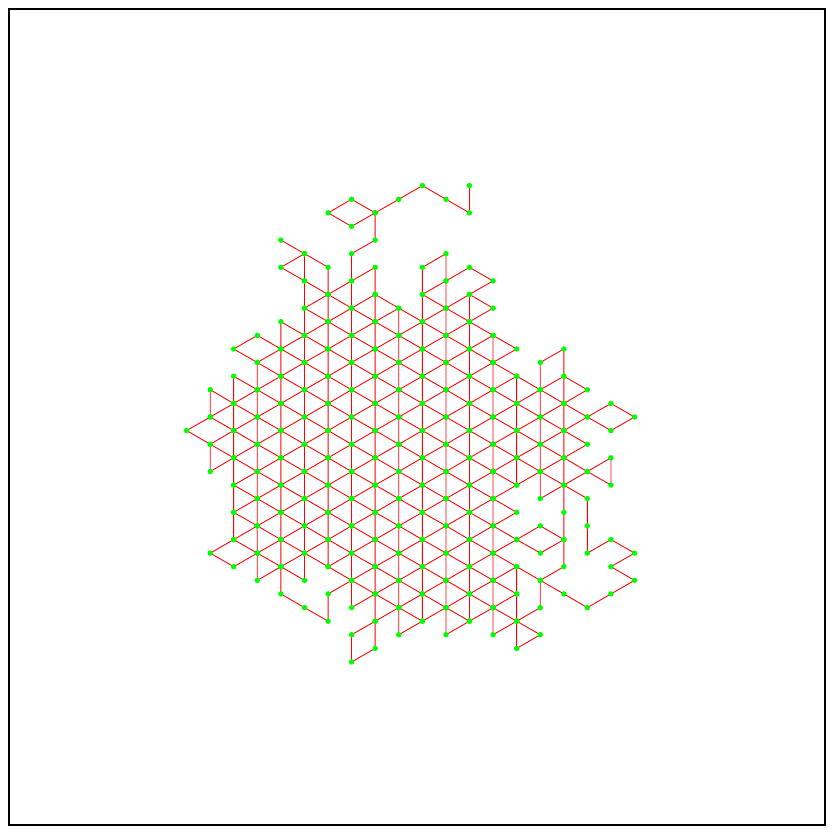}
\end{center}
\caption{A typical configuration at length $512$ produced at $(\omega_2,\omega_3)=(1,10)$ which looks like an ordered crystal.}
\label{densetriplevisits} 
\end{figure}

To test the possibility that the low temperature collapsed phase when $k>k_G$ is frozen we have plotted in Figure~\ref{plot_density_k_6} the proportion of steps of the trail \emph{not} involved with a triply-visited site. Given the collapsed nature of the phase it is appropriate to assume that the polymer has a surface \cite{owczarek1993c-:a} and so we plot our estimate against $1/n^{1/2}$. 
We have chosen a point in the $k=6$ model. The quantity $[1-3 u_3(n)]$ is tending to zero as $n$ is increased, suggesting that, proportionally, all sites  of the trail will be triply-visited: we can deduce that configurations produced are maximally dense and solid-like. In Figure~\ref{densetriplevisits} we display a typical configuration produced by our simulations when $(\omega_2,\omega_3)=(1,10)$: it is space-filling, forming a crystal-like structure, being almost totally made up of triply-visited sites.
We find the same behaviour for the $k=k_G$ model at low temperatures.

\begin{figure}[ht!]
\begin{center}
\includegraphics[width=0.9\columnwidth]{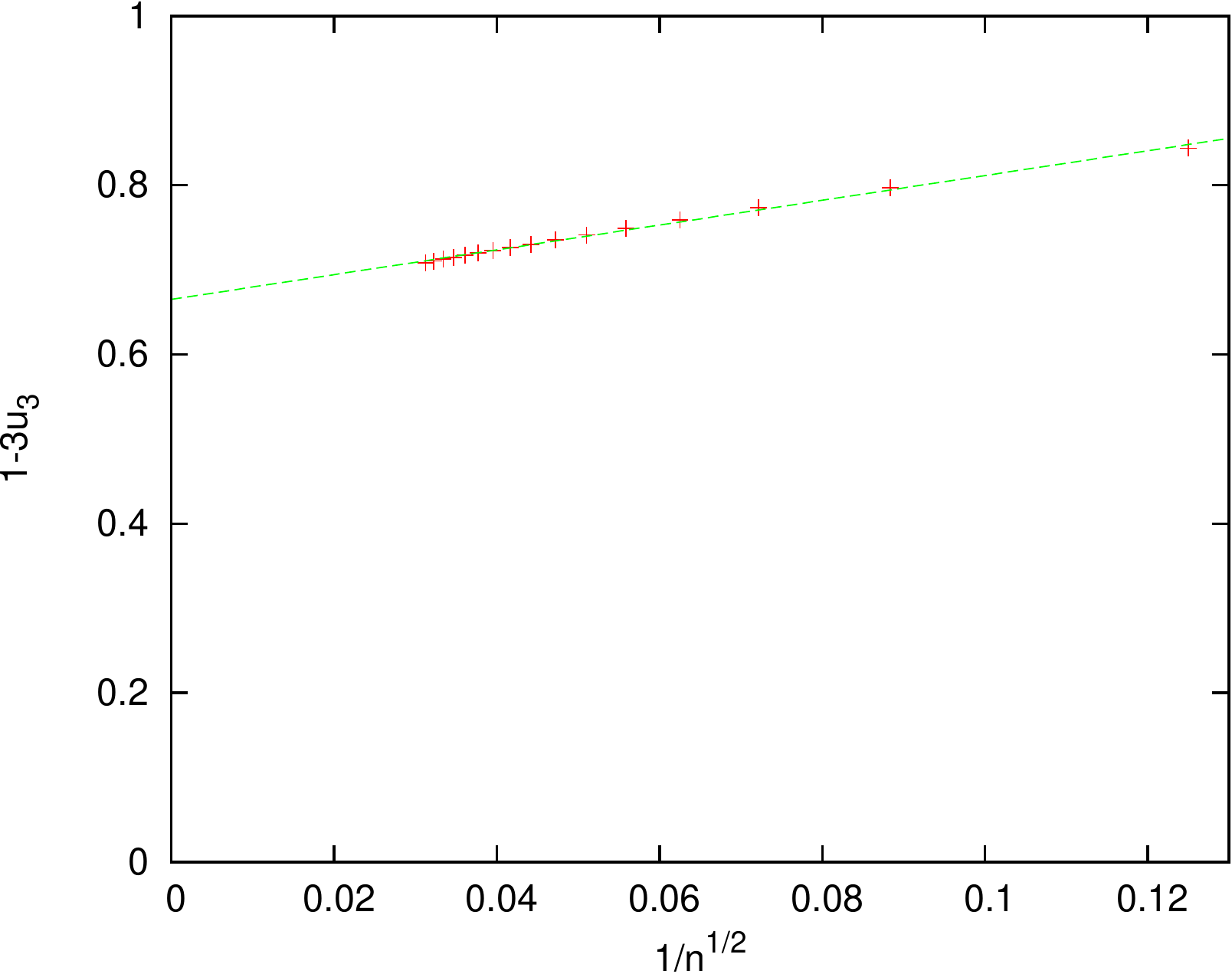}
\end{center}
\caption{Plot of $1-3 u_3(n)$, which measures the proportion of steps that  are not involved with triply-visited sites per unit length, against $1/\sqrt{n}$ at a point  $(\omega_2,\omega_3)=(4,16)$ in the collapsed liquid-drop-like globule phase. As the length increases this reaches a non-zero value.}
\label{density_vs_n} 
\end{figure}

\begin{figure}
\begin{center}
\includegraphics[width=0.7\columnwidth]{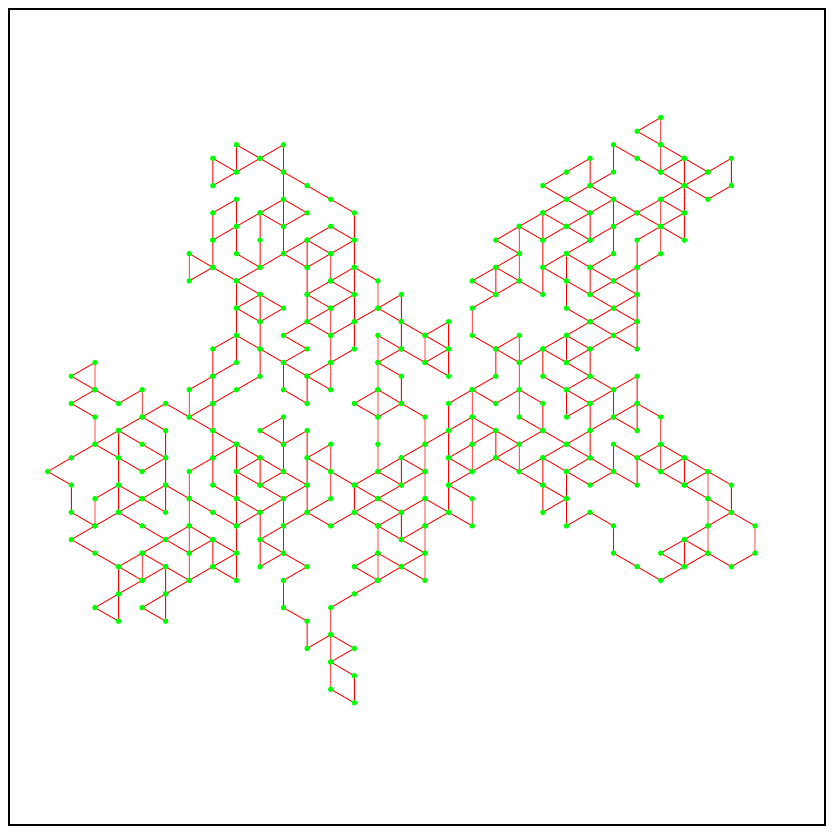}
\end{center}
\caption{ A typical configuration at length $512$ produced at $(\omega_2,\omega_3)=(5,1)$, which is in the globule phase: it looks disordered and rather more like a liquid-like globule than a crystal.}
\label{densedoublevisits} 
\end{figure}

For the sake of comparison we have considered the same quantity $1-3 u_3(n)$ in the low temperature region of the $k=2$ model. In Figure~\ref{density_vs_n} we plot the proportion of steps that  are not involved with triply-visited sites per unit length against $1/n^{1/2}$ at the point $(\omega_2,\omega_3)=(4,16)$: this quantity clearly converges to a non-zero value. 
 In Figure~\ref{densedoublevisits} we display a typical configuration produced by our simulations when $(\omega_2,\omega_3)=(5,1)$: while dense, the configuration still has internal holes and seems disordered. 

From the above considerations it seems that the phases that exist in our eISAT model on the triangular lattice are of a similar type to those in the semi-flexible ISAW phase diagram \cite{bastolla1997a-a,krawczyk2009a-:a,krawczyk2010a-:a}: namely a high temperature swollen phase and two low temperatures phases: one a disordered liquid drop-like globular phase and a second crystalline-like phase. 

\section{Conclusions}

\begin{figure}[hbt]
\begin{center}
\includegraphics[width=0.9\columnwidth]{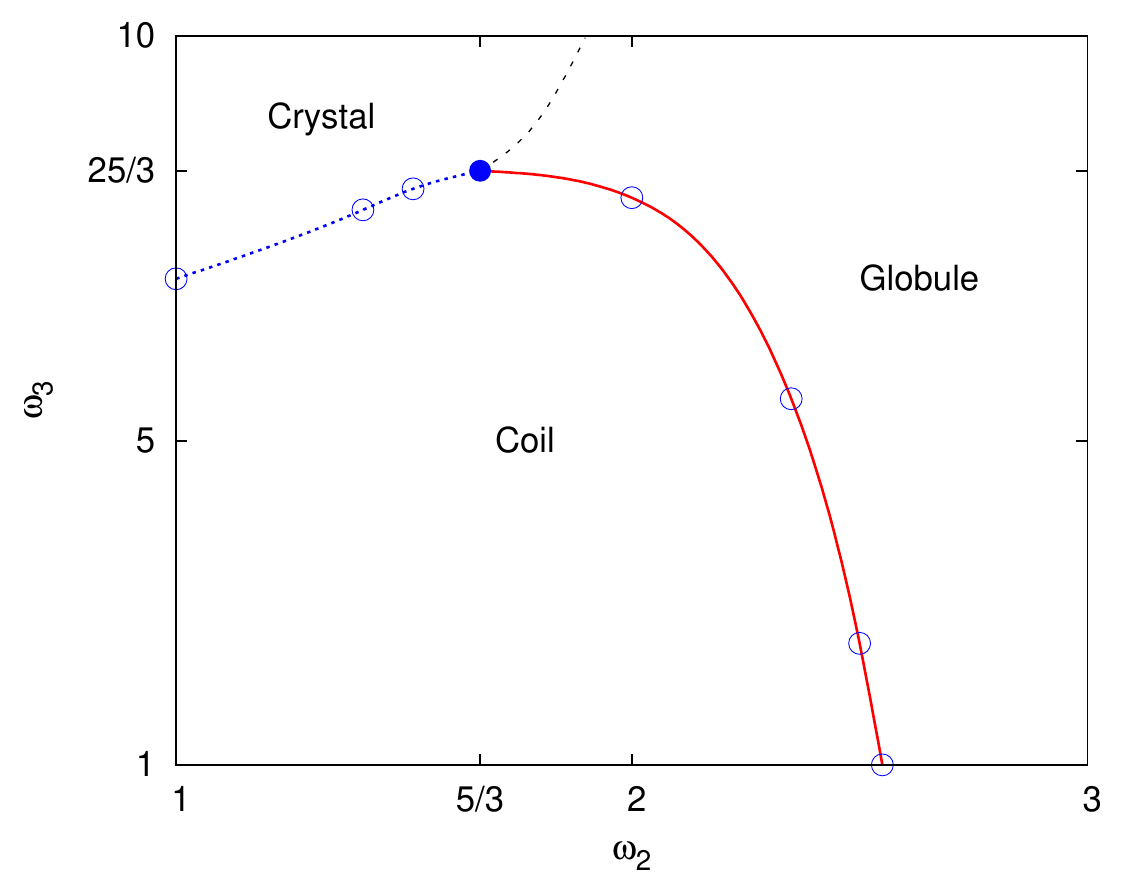}
\end{center}
\caption{Schematic of the proposed phase diagram of the extended ISAT model on the triangular lattice. The filled circle is at the location of the kinetic growth point, and the open circles represent estimates of the
collapse transition for various values of $k$.}
\label{phasediagram} 
\end{figure}

We can now build a complete picture of the phase diagram of the extended ISAT model on the triangular lattice. A schematic of the conjectured phase diagram is shown in Figure~\ref{phasediagram}.
For small $\omega_2$ and $\omega_3$ we see the usual swollen polymer phase where $\nu=3/4$. For large enough $\omega_2$ regardless of $\omega_3$ we find a collapse phase as occurs in the ISAW model and a transition between the swollen and collapsed globule phases which seems to be $\theta$-like. On the other hand for large enough $\omega_3$ we find crystal-like configurations that are space filling and internally contain only triply-visited sites. Between the swollen phase and the crystal-like phase the collapse transition is first-order. Separating this line of first-order transitions from the line of $\theta$-like transitions is a multi-critical point. This point is precisely the point to which the kinetic growth process of trails maps. 

It would seem that this larger parameter space has exposed a way of understanding the apparent differences and similarities of the ISAT and ISAW models. When ISAW is generalised with the addition of stiffness, and ISAT on a large enough coordination number lattice  is generalised with different weightings for different numbers of visits, both display three polymer phases: swollen coil, globule and polymer crystal. Note that for the eISAT model considered here, the crystal-like phase has non-zero entropy and as such is strictly speaking not a proper crystalline phase, but rather a maximally dense phase. This is in contrast to the polymer crystal phase found in the semi-flexible ISAW model \cite{bastolla1997a-a,krawczyk2009a-:a,krawczyk2010a-:a}, where one finds zero-entropy $\beta$-sheet like structures.

In two dimensions both semi-flexible ISAW and eISAT have a first-order collapse transition between swollen coil and crystal-like phases, and a $\theta$-point like transition between swollen coil and globule phases. It is also apparent that the square lattice ISAT model is unusual in that it only displays the multi-critical point, which is found  in these generalised models as part of a larger phase diagram. 

\vspace{2em}

\begin{acknowledgments} 
 Financial support from the Australian
Research Council via its support for the Centre of Excellence for Mathematics and Statistics
of Complex Systems is gratefully acknowledged by the authors. J Doukas was in part supported
by the Japan Society for the Promotion of Science (JSPS), under fellowship no.\ P09749. The
FlatPERM simulations were performed on the computational resources of the Victorian Partnership for
Advanced Computing (VPAC).
A L Owczarek thanks the School of Mathematical Sciences, Queen Mary, University of London for hospitality.
\end{acknowledgments}

\end{document}